\documentclass[aps,twocolumn,showpacs,floatfix]{revtex4}

\usepackage{dcolumn,multirow}
\usepackage{graphicx}
\usepackage{amsfonts,mathtools,amsmath,bm,amssymb}
\usepackage{comment}
\usepackage[usenames]{color}
\usepackage[flushleft]{threeparttable}
\usepackage{titlesec}
\usepackage{etoolbox} 
\usepackage{lipsum} 
\usepackage[capitalize]{cleveref}

\usepackage{color}

\begin{document}

\title{ Magnetism of (LaCoO$_3$)$_n$+(LaTiO$_3$)$_n$ superlattices with $n=1,2$}

\author{Alex Taekyung Lee}
\affiliation{Department of Applied Physics, Yale University }

\author{Sohrab Ismail-Beigi}
\affiliation{Department of Applied Physics, Yale University }

\date{\today }

\begin{abstract}

LaCoO$_3$ provides a poignant example of a transition metal oxide where the cobalt cations display multiple spin states and spin transitions and which continues to garner substantial attention.  In this work, we describe first principles studies, based on DFT+$U$ theory, of superlattices containing LaCoO$_3$, specifically (LaCoO$_3$)$_n$+(LaTiO$_3$)$_n$ for $n=1,2$.  The superlattices show strong electron transfer from Ti to Co resulting in Co$^{2+}$, significant structural distortions and a robust orbital polarization of Co$^{2+}$. We predict high-spin Co$^{2+}$ and a checkerboard  or G-type antiferromagnetic (AFM) ground state.  We provide a detailed analysis of the magnetic interactions and phases in the superlattices.  We predict that ferromagnetic order on the Co${2+}$ can be stabilized by hole doping (e.g., replacing La by Sr) which is rather unusual for Co$^{2+}$ cations.

\end{abstract}

\pacs{75.70.Cn, 73.20.-r, 75.47.Lx, 71.15.Mb }

\maketitle

\section{Introduction}

Transition metal perovskite oxides (with chemical formula $AB$O$_3$, where $A$ is a rare earth and $B$ is a transition metal) display fascinating physical properties,
such as high-temperature superconductivity in the cuprates or colossal magnetoresistance in the manganites \cite{LeeRMP2006,Tokura2000}. The electronic and structural properties of perovskites and intertwined and lead to complex behaviors due to the interplay of charge, magnetic, structural, and orbital degrees of freedom \cite{Hwang2012}.  Creating heterostructures of transition metal perovskites enlarges the playground for manipulation of the local structure, charge state and electronic properties of these materials.  In this work, we focus on the electronic states of cobalt-containing perovkites.

Bulk LaCoO$_3$ has attracted much attention
because of the complex electronic and magnetic structure of the Co$^{3+}$ cations it contains.  LaCoO$_3$ is a nonmagnetic insulator at the lowest temperatures \cite{BhidePRB1972,ZobelPRB2002} when the 
Co$^{3+}$ adopt a low-spin (LS) state ($t^{6}_{2g}$, $S$=0). It becomes a paramagnetic insulator for temperatures 100 K $ < T <$ 500 K and then becomes metallic above 500 K. 
However, the spin state of Co in the thermally excited material is still in debate. The magnetic transition near 100 K is  considered as a transition from LS to high-spin (HS)  
($t^{4}_{2g}e^{2}_{g}$, $S=2$) \cite{AsaiPRB1994,ItohJPSJ1994}
or to intermediate spin (IS)  ($t^{5}_{2g}e^{1}_{g}$, $S=1$) states
\cite{PotzePRB1995,SaitohPRB1997,AsaiJPSJ1998,YamaguchiPRB1997,ZobelPRB2002},
while IS and HS are mixed as temperature increases \cite{YamaguchiPRB1997,AsaiJPSJ1998}.

Numerous first principles calculations based on DFT+$U$ theory \cite{LDA+U1} for LaCoO$_3$ confirm that the LS state is the ground state 
\cite{KorotinPRB1996,KnizekPRB2005,KnizekJPCM2006}, while the IS state is more stable than HS \cite{KorotinPRB1996}.
On the other hand, several theoretical studies suggested that the excited state at 100 K $< T <$ 500 K can be a mixture of HS and LS,  which we deonte as HS/LS below. Using the experimental crystal structures as a function of temperature, GGA+$U$ calculations (with $U=$ 2.7 eV) calculations suggested that HS/LS is stabilized for the structures above 200 K \cite{KnizekJPCM2006}. Unrestricted Hartree-Fock calculations also proposed that the non-magnetic to paramagnetic transition 
is due to the LS state to LS/HS state \cite{ZhuangPRB1998}, 
and  effective Hamiltonian calculations suggested a similar scenario \cite{RopkaPRB2003,BiernackiPRB2006}.
Recent DFT+dynamical mean-field theory calculations showed that LaCoO$_3$ has large charge fluctuations: these calculations show that it is not possible to explain the spin state with a single multiplet at any temperature, although they find that at the onset of the spin-spin transition
HS multiplets are excited with the IS multiplets being excited later around the onset of the metal-insulator transition.

On the other hand, the electronic and magnetic structures of Co$^{2+}$ in bulk CoO does not show such complexity or ambiguity.
At low temperatures, CoO is an insulator with energy gap $E_g =2.4$ eV \cite{ZimmermannJPCM1999}.
The Co$^{2+}$ cations have a HS state ($t^{5}_{2g}e^{2}_{g}$ with $S=3/2$), and  antiferromagnetically  order at $T_N=$  289 K with a magnetic ordering wave vector of $\bm{q_\textnormal{fcc}} = \left(\frac{1}{2},\frac{1}{2},\frac{1}{2}\right)$ 
\cite{GreenwaldAC1953,TomiyasuPRB2004}.
Unlike LaCoO$_3$, Co$^{2+}$ in CoO does not show LS or IS states nor a spin-state transition. 

An intruiguing set of questions arise as to what happens if the Co$^{2+}$ cations are realized in a perovskite heterostructure containing LaCoO$_3$.  Is the difference between the electronic behavior of LaCoO$_3$ and CoO only due to charge state of the Co cations?  To what extent can the Co electronic properties be controlled or manipulated by changing the local lattice structure surrounding the Co cation?  And how can one made strong modifications to the local structure?  One way to control the charge state of transition metal cations and the local lattice structure is via heterostructuring. For example, charge transfer can be triggered from Ti to Ni in LaNiO$_3$+LaTiO$_3$ superlattices~\cite{ChenPRL2013,DisaPRL2015} which results in a  Ni$^{2+}$ charge state with large $e_g$ orbital polarization,
which is in strong contrast to the orbitally unpolarized Ni$^{3+}$ in bulk LaNiO$_3$.  Recent work on LaCoO$_3$+LaTiO$_3$ superlattices show similar charge transfer and orbital polarization as well~\cite{our PRL once it is out soon}, which we examine in great detail here theoretically \cite{JaePRL2019}.

In this work, we study the electronic and magnetic properties of Co in (LaTiO$_3$)$_1$+(LaCoO$_3$)$_1$
and (LaTiO$_3$)$_2$+(LaCoO$_3$)$_2$ superlattices using DFT+$U$ theory.  We explore carefully the effect of the Hubbard $U$ parameter, account for the effects of different exchange-correlation approximations, examine the effect of strain from the substrate, as well as the effects of doping.
We explain the origin of the magnetic stabilities for different spin states and magnetic orderings by using a simple band interaction picture.  Interestingly, from the analysis of the magnetic stabilities, we are able to show that ferromagnetic spin order on for the Co$^{2+}$ cations can be stabilized by the hole doping which is an unexpected and unique finding for Co$^{2+}$ as detailed below (Summary and outlook section).

This paper is organized as follows.
In Sec. \ref{sec:bulk}, we study the $U_{\textnormal{Co}}$-dependent properties of bulk LaCoO$_3$ and CoO,
which have Co$^{3+}$ and Co$^{2+}$ cations, respectively; this allows us to find ranges of  $U_{\textnormal{Co}}$ that describe key electronic and magnetic properties of two materials correctly when compared to experiment.
In Sec. \ref{sec:general_aspects}, we describe the general atomic, electronic, and magnetic properties of 
LaTiO$_3$+LaCoO$_3$ superlattices.
In Sec. \ref{sec:LTO1+LCO1}, we discuss the electronic and magnetic properties of 
(LaCoO$_3$)$_1$+(LaTiO$_3$)$_1$ superlattices.
In Sec. \ref{sec:spin_states}, we describe the energetics of HS and LS states 
as a function of $U_{\textnormal{Co}}$ and examine the origin of the $U_{\textnormal{Co}}$ dependences.
We present the magnetic stabilities of Co as a function of $U_{\textnormal{Co}}$ in Sec.~\ref{sec:mag_stability}
and explain using a simple band interaction model in Sec.~\ref{sec:mag_origin}.
Effects of strain and different exchange-correlation approximations are presented in Sec.~\ref{sec:strain}.
In Sec.~\ref{sec:doping}, we show that the ferromagnetic phase is stabilized by hole-doping in 
(LaCoO$_3$)$_1$+(LaTiO$_3$)$_1$, and we explain the nature of its magnetic stability.  In Sec.~\ref{sec:LTO2+LCO2}, we repeat our analysis of electronic and magnetic properties for 
(LaCoO$_3$)$_2$+(LaTiO$_3$)$_2$ and examine the importance of out-of-plane Co--Co magnetic interaction.

\section{Computational Details}

We use density functional theory (DFT) with the projector augmented wave (PAW) method \cite{PAW} and the revised version of the generalized gradient
approximation (GGA) proposed by Perdew {\it et al.} (PBEsol) \cite{PBEsol} as implemented in the VASP software \cite{VASP}.
We also have done selective tests using the local density approximation (LDA) \cite{CA,Perdew}.
In all cases, the spin-dependent version of the exchange correlation functional is employed.
A plane wave basis with a kinetic energy cutoff of 500 eV is used. 
We use $\Gamma$-centered \textbf{k}-point meshes of size  9$\times$9$\times$7 and 9$\times$9$\times$5 
for (LaCoO$_3$)$_1$+(LaTiO$_3$)$_1$ and (LaCoO$_3$)$_2$+(LaTiO$_3$)$_2$, respectively. 
The size of the supercells for (LaCoO$_3$)$_1$+(LaTiO$_3$)$_1$ and (LaCoO$_3$)$_2$+(LaTiO$_3$)$_2$
are 20 and 40 atoms, respectively, correspond to the $\sqrt 2 \times \sqrt2$ (i.e., $c(2\times2)$) in-plane unit cells.
These unit cells have two Co atoms in each CoO$_2$ plane, so that the 
(LaCoO$_3$)$_1$+(LaTiO$_3$)$_1$ and (LaCoO$_3$)$_2$+(LaTiO$_3$)$_2$ each have 2 and 4 Co atoms in their unit cells, respectively.
Atomic positions within the
unit cells were relaxed until the residual forces were less than 0.01 eV/\AA{} in all axial directions, and the stress is less than 0.5kB along the $z$ axis.

The GGA+$U$ scheme within the rotationally invariant formalism together with the fully localized
limit double-counting formula \cite{LDA+U1} is used to study the effect of electron interactions.  
The electronic and structural properties critically depend on
$U_{\textnormal{Co}}$, and  therefore we carefully explore a range of values.  
We also explore how the results depend on $U_{\textnormal{Ti}}$, 
which plays a secondary role in the physics of these materials
We do not employ an on-site exchange interaction $J$ for any species, as the exchange interaction is already accounted for within the spin-dependent DFT exchange-correlation potential \cite{Hyowon,Chen2015}.    Finally, we used Maximally Localized Wannier Functions (MWLF) as implemented in the Wannier90 software~\cite{wannier90} to obtain the localized tight-binding representation describing Co 3d orbitals and their hybridizaiton.

\subsection{$U$ dependence of LaCoO$_3$ \& CoO}
\label{sec:bulk}

\begin{figure}
\begin{center}
\includegraphics[width=0.5\textwidth, angle=0]{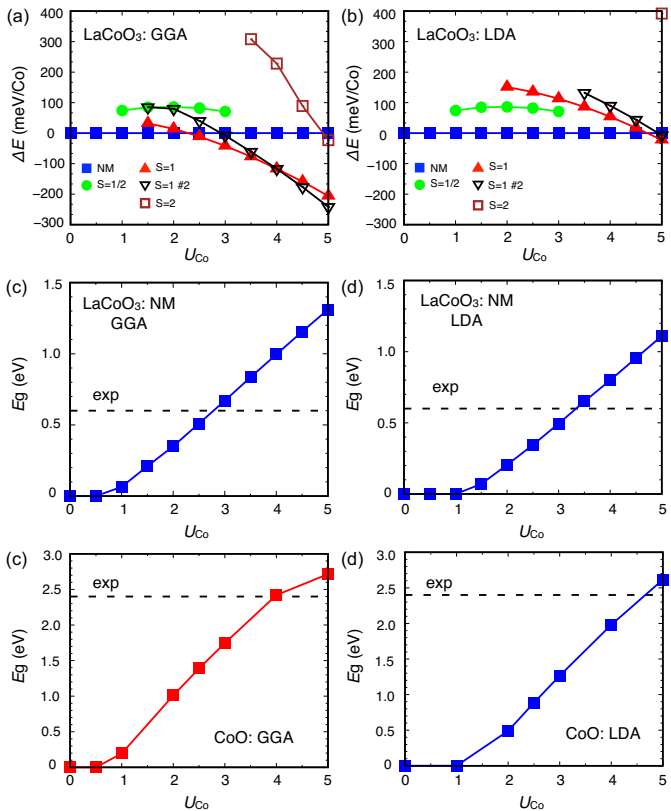}
\caption{Total energies of different magnetic states of bulk LaCoO$_3$ within (a) GGA+$U$ and (b) LDA+$U$, and the
band gap of bulk LaCoO$_3$ within (c) GGA+$U$ and (d) LDA+$U$.
Band gap of bulk CoO within (e) GGA+$U$ and (f) LDA+$U$.  The total energies of the nonmagnetic insulating phase (NM) are set to zero.}
\label{bulk}
\end{center}
\end{figure}

In this section, we aim to obtain optimal $U_{\textnormal{Co}}$ value for Co$^{2+}$ in the LCO+LTO superlattices by studying bulk LaCoO$_3$ (Co$^{3+}$) and CoO (Co$^{2+}$).
Below, we will show  that the experimental nonmagnetic ground state for LaCoO$_3$ is stable with $U_{\textnormal{Co}} \leq 2.5$ eV for LaCoO$_3$, while $U_{\textnormal{Co}} \sim 4.5$ eV is needed to obtain the experimental band gap of bulk CoO (see Figure~\ref{bulk}).
We will conclude that $U_{\textnormal{Co}}=3$ eV is a reasonable value.

\subsubsection{Bulk LaCoO$_3$ (LCO)}

We consider the energetics of different magnetic configurations of Co$^{3+}$ in LCO. 
We begin with bulk LaCoO$_3$ within both GGA+$U$ and LDA+$U$ for $0 \leq U_{\textnormal{Co}} \leq 5$ eV 
with the $R\bar{3}c$ crystal structure, with a unit cell containing two Co atoms.
The atomic structures and lattice vectors are  fully relaxed. 
As presented in Figs.~\ref{bulk} (a,b) and Table~\ref{lcotable}, we can stabilize five different Co$^{3+}$spin states:
$S$=0, 1/2, 1, 1\#2, and 2. 
Interestingly, $S$=1/2, 1, and 1\#2 show charge, spin, and bond length disproportionation (see Table~\ref{lcotable}),
while their  structural symmetry is still $R\bar{3}c$.
For example, for $S$=1, the two Co atoms (Co1 and Co2) have different total numbers of $d$ electrons ($N_d$) and Co magnetic moments ($M$),
in addition to different Co--O bond lengths $d_{\textnormal{Co--O}}$:
$N_{d}({\textnormal{Co1}})=7.349$ with $d_{\textnormal{Co1--O}}= 1.90$ \AA{} while  
$N_{d}({\textnormal{Co2}})=7.047$ with $d_{\textnormal{Co2--O}}= 1.98$ \AA.

\begin{table}
\begin{ruledtabular}
\begin{center}
\caption{Magnetic moment $M$ ($\mu_B$) and $N_{d}$  of Co1 and Co2 in bulk LaCoO$_3$. 
$U_{\textnormal{Co}}=2.5$ eV is used, except for the case $S=2$ since this state is not even metastable when $U_{\textnormal{Co}} < 3.5 $ eV. $M$ and $N_d$ are computed from the VASP-calculated local 3d occupancies around each atom.}
\label{lcotable}
\renewcommand{\arraystretch}{1.3}
\begin{tabular}{c c c c c}
& \multicolumn{2}{c}{$M$(Co)} ($\mu_B$))	& \multicolumn{2}{c}{$N_d$(Co)} \\
\cline{2-3} \cline{4-5}
& Co1 & Co2 & Co1 & Co2 \\
 \hline
(i) $S$=0 (NM)		& 0		& 0		& 7.332	& 7.332 \\	
(ii) $S$=1/2	& 0.344	& 1.795 	& 7.375	& 7.219 \\
(iii) $S$=1		& 0.366	& 2.922	& 7.349	& 7.047 \\
(iv) $S$=1\#2	& 1.545	& 2.322	& 7.222	& 7.196 \\
(v)  $S$=2 	& 3.224	& 3.224	& 6.956	& 6.956 \\
\end{tabular}
\end{center}
\end{ruledtabular}
\end{table}

Within GGA+$U$  (Fig. \ref{bulk}(a)),
the ground state of bulk LaCoO$_3$ within GGA+$U$ is $S=0$ (nonmagnetic state, NM) 
for $0 \leq U_{\textnormal{Co}} < 2.5$ eV,
$S=1$ for $2.5 \leq U_{\textnormal{Co}} < 4$ eV,
and $S=1\#2$ for $4 \leq U_{\textnormal{Co}} \leq 5$ eV.  
Thus, within GGA+$U$, $0 \leq U_{\textnormal{Co}} < 2.5$ eV is appropriate to reproduce the nonmagnetic low-temperature ground state in the experiment. Within LDA+$U$, the $S=0$ NM state is the ground state for $0 \leq U_{\textnormal{Co}} \leq 4.5$ eV.

In addition to the energetics of the various spin states, we also consider the band gap ($E_g$) of bulk LaCoO$_3$.
As summarized in Fig.~\ref{bulk} (c), $E_g$ within GGA+$U$ with $U=2.5$ eV and $U=3$ eV are 0.51 and 0.67 eV, respectively, 
which are both close to the the experimental gap of $\approx0.6$ eV \cite{ChainaniPRB1992}.
Within LDA+$U$, $U=3.5$ eV gives $E_g=0.65$ eV, as shown in Fig.~\ref{bulk} (d). 

Considering both the nature of the magnetic ground state and a reasonable value of the energy gap, 
we conclude that $U=2.5$ eV within GGA+$U$ and $U=3.5$ eV within LDA+$U$ are  reasonable values for studying bulk LaCoO$_3$.
Our $U$ values are comparable to those from previous DFT+$U$ studies:
$U_{\textnormal{Co}}-J_{\textnormal{Co}}$= 2.7, 3.5, and 4.5 eV~ \cite{KnizekJPCM2006} 
or LDA+$U$ values of $U_{\textnormal{Co}}=7.8$ eV and $J_{\textnormal{Co}}= 0.92$ eV~\cite{KorotinPRB1996}.

\subsubsection{Bulk CoO}
In the LaCoO$_3$+LaTiO$_3$ superlattice, the electron transfer from Ti to Co will lead to a 2+ charge state for Co (see the next section).  
Therefore, we also calculate the $U_{\textnormal{Co}}$ dependence of the band gap of bulk CoO which also contains Co$^{2+}$.

At low temperature, bulk CoO is antiferromagnetic with wave vector 
$\bm{q_\textnormal{fcc}} = \left(\frac{1}{2},\frac{1}{2},\frac{1}{2}\right)$ 
\cite{GreenwaldAC1953,TomiyasuPRB2004},
and an insulator with energy gap $E_g =2.4$ eV \cite{ZimmermannJPCM1999}.
In Figs. \ref{bulk} (c) and (d), we plot $E_g$ of CoO as a function of $U_{\textnormal{Co}}$.
$U_{\textnormal{Co}}=4$ eV within GGA+$U$ and $U_{\textnormal{Co}}\sim4.8$ eV within LDA+$U$ 
give the experimental gap. 
These values are similar to the previous GGA+$U$ results: 
$E_g =2$ eV with $U_{\textnormal{Co}}-J_{\textnormal{Co}}$= 3.3 eV  \cite{WangPRB2006}, 
$E_g =2.8$ eV with $U_{\textnormal{Co}}=7.1$ eV and $J_{\textnormal{Co}}= 1$ eV \cite{WangPRB2006}.

\subsubsection{Choosing $U_{\textnormal{Co}}$}

Within GGA+$U$, we need $U_{\textnormal{Co}} = 2.5$ eV for bulk LaCoO$_3$ to be described well, and $U_{\textnormal{Co}}=4$ eV for bulk CoO.   Since we expect out superlattice to contain Co$^{2+}$ which is more similar to CoO, we choose $U_{\textnormal{Co}}=3$ eV  unless otherwise specified.
Below, we will explore $0 \leq U_{\textnormal{Co}} \leq 5$ eV to study the $U$-dependence of the electronic and magnetic properties.
We consider $U_{\textnormal{Ti}}$= 0, 3, and 5 eV, and use $U_{\textnormal{Ti}}=3$ eV unless specified otherwise.
This value of $U_{\textnormal{Ti}}$ is obtained by fitting the optical gap of LaTiO$_3$ to be 0.2 eV based on our previous work \cite{LTOLAO}.

\section{ LaCoO$_3$+LaTiO$_3$ superlattices}
\label{sec:LCO+LTO}

\subsection{General aspects}
\label{sec:general_aspects}

In this subsection, we discuss the overall atomic, electronic, and magnetic properties of LaCoO$_3$+LaTiO$_3$ superlattices.
We will show that an electron is transferred from Ti to Co resulting in Ti$^{4+}$ ($d^0$) and Co$^{2+}$ ($d^7$).
This leads to a local electric field between CoO$_2$ and TiO$_2$ layers which results in large shifts of the La$^{3+}$ and O$^{2-}$ ions between them and an elongation of interfacial Co--O bond.  Electronically, the main observation of this section is that the 
the interfacial Co$^{2+}$ can have a high-spin (HS, $t^{5}_{2g}e^{2}_{g}$) or low-spin state (LS, $t^{6}_{2g}e^{1}_{g}$).

\begin{figure}
\begin{center}
\includegraphics[width=0.5\textwidth, angle=0]{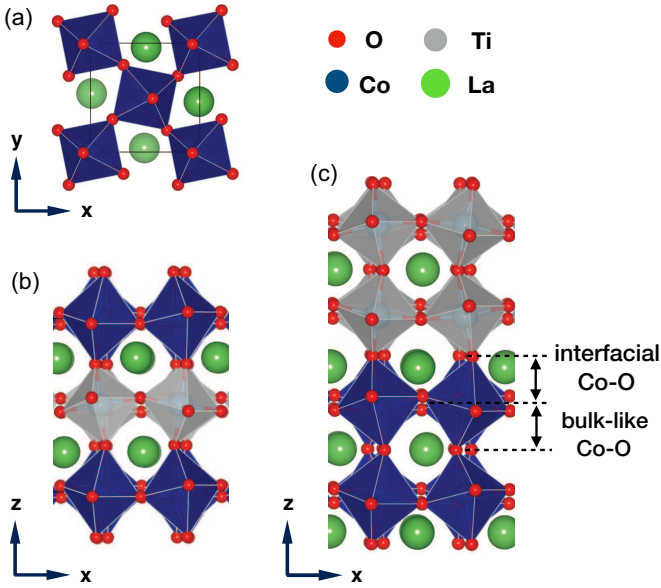}
\caption{Atomic-scale structures of (LaCoO$_3$)$_1$+(LaTiO$_3$)$_1$ and (LaCoO$_3$)$_2$+(LaTiO$_3$)$_2$ superlattices. 
(a) Top view and  (b) side view of the (LaCoO$_3$)$_1$+(LaTiO$_3$)$_1$ superlattice.  
(c) Side view of the (LaCoO$_3$)$_2$+(LaTiO$_3$)$_2$ superlatttice.
}
\label{atm_str}
\end{center}
\end{figure}

We begin by discussing the structure of the LaCoO$_3$+LaTiO$_3$ (LCO+LTO) superlattices. In addition to the relaxation of the shape of the local oxygen octahedra, resulting in Jahn-Teller-like distortions, tilting of oxygen octahedra is also very important in determining the electronic structure of  transition metal oxides. Unlike bulk oxides, since the tilt pattern of the superlattice is harder to observe (due to thin samples) and is not known {\it a priori}, we consider several tilt patterns and choose the most stable one. Since bulk LCO has the $R\bar{3}c$ crystal structure with $a^-a^-a^-$ octahedral tilts and bulk LaTiO$_3$ has the $Pbnm$ structure with $a^-a^-b^+$ tilts, 
we examine three cases: $a^0a^0a^0$, $a^-a^-a^-$ and $a^-a^-b^+$.
We find that $a^-a^-b^+$ is always the most stable, so we only examine this tilt pattern below unless further specified. 

The structures are relaxed along the $c$ axis, which is normal to the interface, while in-plane lattice parameters $a$ and $b$ are fixed and set equal $a=b$ to simulate epitaxial strain from a substrate. Three sets of in-plane lattice parameters are used: 3.663 \AA, 3.784 \AA, and 3.905 \AA: 3.784 \AA\ and 3.905 \AA\  are the experimental lattice parameters of LaAlO$_3$ (LAO) and SrTiO$_3$ (STO) substrates, respectively. 
The optimized lattice parameter of bulk LCO within GGA+$U$ ($U_{\textnormal{Co}}=3$ eV) is 3.805 \AA ,close to the LAO lattice.
To study the effect of the compressive strain on LCO, we use 3.663 \AA\ which is smaller than LCO lattice.

\begin{figure}
\begin{center}
\includegraphics[width=0.5\textwidth, angle=0]{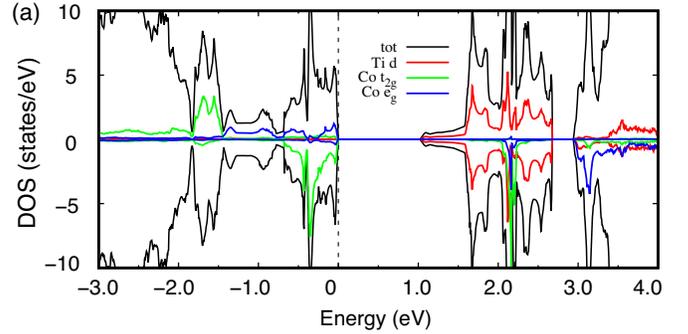}
\caption{Total (black) and projected (colors) density of states (DOS) of (LaCoO$_3$)$_1$+(LaTiO$_3$)$_1$ superlattice onto one Co atom.
The Co atoms have high-spin states and AFM spin alignment.
$U_{\textnormal{Co}}=3$ eV and $U_{\textnormal{Ti}}=3$ eV with GGA+$U$ are used, and the in-plane lattice parameter is $a=3.784$ \AA. 
Positive and negative DOS describe spin up and spin down electronic states, respectively. 
 }
\end{center}
\label{tdos}
\end{figure}

\begin{figure}
\begin{center}
\includegraphics[width=0.5\textwidth, angle=0]{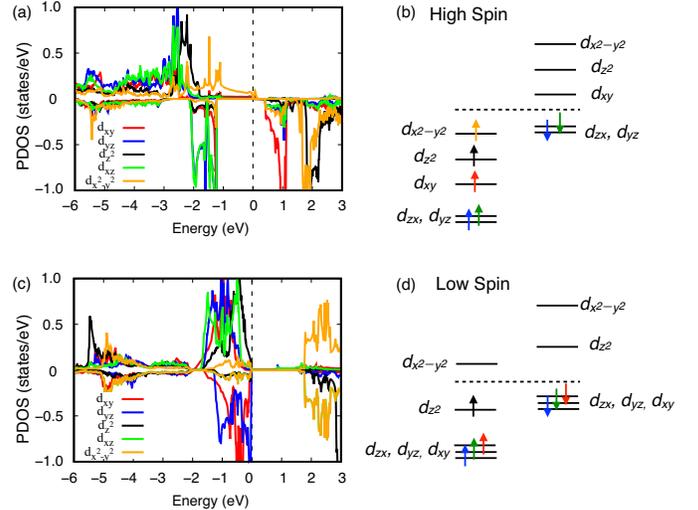}
\caption{Projected density of states of (LaCoO$_3$)$_1$+(LaTiO$_3$)$_1$ for (a) high-spin and (c) low-spin FM states.  
(b) and (d) show schematics of the corresponding atomic-like energy levels. 
$U_{\textnormal{Co}}=3$ eV and $U_{\textnormal{Ti}}=3$ eV with GGA+$U$ are used, and the in-plane lattice parameter is $a=3.784$ \AA. 
}
\label{HS_LS_pdos}
\end{center}
\end{figure}

Next, we discuss the charge states of transition metals and the electron transfer.
Given the difference in electronegativity between Ti and Co \cite{ALLRED1961215,AllenJACS1989}, 
 electron transfer is expected from Ti 3$d$ to Co 3$d$ across the interface. 
Projected densities of states (Fig.~\ref{tdos}) within GGA+$U$ show that the Ti $d$ bands are completely empty, which corresponds to Ti$^{4+}$, and describe the conduction band minimum.
On the other hand, Co has 2+ charge state with $d^7$ electrons, indicating full charge transfer.

\begin{figure}
\begin{center}
\includegraphics[width=0.45\textwidth, angle=0]{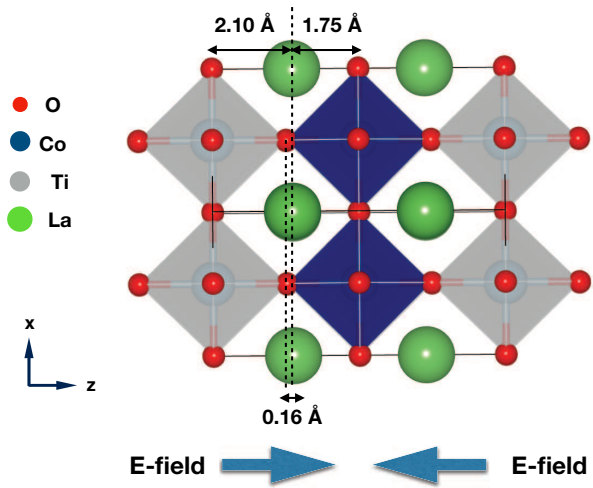}
\caption{Atomic structure of the (LaCoO$_3$)$_1$+(LaTiO$_3$)$_1$ superlattice with $a^0a^0a^0$ octahderal tilt
and schematic picture of the local electric field.
Distances between the La plane and the TiO$_2$ or CoO$_2$ planeas as well as the shift between La and O in the same plane (prior to displacement) are shown on the figure.  The structure is obtained by using $U_{\textnormal{Co}}=3$ eV, $U_{\textnormal{Ti}}=2.5$ eV, GGA+$U$, and an in-plane lattice parameter of $a=3.784$ \AA. 
  }
\label{Efield}
\end{center}
\end{figure}

Due to the charge transfer, we have Ti$^{4+}$ and Co$^{2+}$ and we expect local electric fields going from the TiO$2$ layer to the CoO$_2$ layer.  This can be confirmed by looking at the positions of the positively charged La cations (La$^{2+}$) and negatively charged apical oxygen anions (O$^{2-}$): the displacements are  visible to the eye in Fig.~\ref{atm_str}.  For example, for 
 (LCO)$_2$+(LTO)$_2$ on an STO substrate, the out of plane Co--O bond length between the CoO$_2$ planes (denoted as bulk-like Co--O bond in Fig.~\ref{atm_str}(c)) is 2.02 \AA, 
while the bond length between the CoO$_2$ and TiO$_2$ planes (denoted as interfacial Co--O bond in Fig.~\ref{atm_str}(c)) is 2.28 \AA, indicating that O$^{2-}$ is shifted to the opposite direction of the local electric field.  A simpler theoretical view unencumbered by octahedral tilting is afforded by considering (LCO)$_1$+(LTO)$_1$ with $a^0a^0a^0$ tilt within GGA+$U$ whose structure is shown in Fig.~\ref{Efield}.  Here, the distance between the La$^{3+}$ plane and the CoO$_2$ plane is 1.75 \AA, while the distance between La$^{3+}$ and TiO$_2$ plane is 2.10 \AA.

We now discuss the Co$^{2+}$ spin states.  
The HS state  ($t^{5}_{2g}e^{2}_{g}$) with $S=1$ is depicted in Figs.~\ref{HS_LS_pdos}(a) and (b): the spin-up $d$ bands are fully occupied, while spin-down $d$ bands have two electrons in the $t_{2g}$ channel.
Due to the non-zero $U_{\textnormal{Co}}$, the spin-down $t_{2g}$ splits into two degenerate bands ($d_{xz},d_{yz}$) under the Fermi level 
and one empty single band ($d_{xy}$) above the Fermi level, resulting in a Mott-like insulator.
We find that $U_{\textnormal{Co}}>0$ is necessary  for opening a energy gap for the HS state.  In addition, the HS state is not even metastable at $U_{\textnormal{Co}}=0$:
$U_{\textnormal{Co}}>2$ eV is needed (with $U_{\textnormal{Ti}}=3$ eV) to obtain a gapped HS state.  Regarding the orbital polarization,  
we find that the degeneracy of the $t_{2g}$ is already removed at $U_{\textnormal{Co}}=0$
because of the broken symmetry due to  interface formation.
This splitting  between $d_{xz}$/$d_{yz}$ and $d_{xy}$ is further enhanced by the imposed epitaxial strain and also enlarged by the non-zero $U_{\textnormal{Co}}$.

The LS state ($t^{6}_{2g}e^{1}_{g}$) has $S=1/2$ and is presented in Figs.~\ref{HS_LS_pdos}(c) and (d):
the $t_{2g}$ bands are fully occupied, while the one remaining electron is in the $e_g$ channel.  The degeneracy of the $e_g$ manifold is already broken at $U_{\textnormal{Co}}=0$ due to the interface formation and epitaxial strain.  
Due to the large degeneracy splitting of the $e_g$ bands, only the spin-up $d_{z^2}$ is filled in the LS state while the spin-up $d_{x^2-y^2}$ and spin-down $e_g$ bands are empty.
As a result, LS state is already insulating at $U_{\textnormal{Co}}=0$ and has strong $e_g$ orbital polarization (and $U_{\textnormal{Co}}>0$ enhances the polarization and the energy gap).

\subsection{(LaCoO$_3$)$_1$+(LaTiO$_3$)$_1$}
\label{sec:LTO1+LCO1}

In this subsection, we discuss, in great detail, the electronic and magnetic properties of the (LaCoO$_3$)$_1$+(LaTiO$_3$)$_1$ superlatice ((LTO)$_1$+(LCO)$_1$ below).
In Sec.~\ref{sec:spin_states}, we analyze the energy difference between high-spin and low-spin states 
as a function of $U_{\textnormal{Co}}$.
In Sec.~\ref{sec:mag_stability}, we examine the $U_{\textnormal{Co}}$ dependence of the the energy difference 
between ferromagnetic and antiferromagnetic spin orders.
We study the effects of strain and different exchange correlation functionals in Sec.~\ref{sec:strain}.

The material system of interest is (LTO)$_1$+(LCO)$_1$ stacked into an infinite superlattice along the [001] direction as depicted in Fig.~\ref{atm_str}(c).
As mentioned above, three in-plane lattice parameters are considered: $a$=3.663, 3.784, and 3.905 \AA. 
We first focus on $a=3.784$ \AA, which corresponds to the LAO substrate, and discuss the effect of the strain later. 
In (LTO)$_1$+(LCO)$_1$, the CoO$_6$ octahedron has  interfaces on both sides.
Therefore, both out-of-plane Co--O bonds are elongated, due to the local electric field  discussed above. 
When $a=3.784$ \AA, the in-plane Co--O bond lengths are 1.947 and 1.950 \AA, whereas the out-of-plane Co--O bond length is 2.205 \AA.

Our supercell of (LTO)$_1$+(LCO)$_1$ has a $c(2\times2)$ in-plane unit cell and therefore two distinct Co atoms, which can be HS or LS separately, and also can be spin-up or spin-down separately.
Therefore, there are 6 different spin configurations: 
ferromagnetic with two high-spin Co  (HS FM), 
antiferromagnetic with two high-spin Co (HS AFM), 
ferromagnetic with one high-spin and one low-spin Co  (HS/LS FM),
ferrimagnetic with one high-spin and one low-spin Co  (HS/LS FIM),
ferromagnetic with two low-spin Co  (LS FM), and 
antiferromagnetic with two low-spin Co (LS AFM).
Among them, HS AFM is the most stable over a wide range of parameters, as shown in Fig.~\ref{HS_LS_Udep}(a).
For fixed spin states, HS AFM is more stable than HS FM (Fig. \ref{mag_Udep}(a)), 
HS/LS FM is more stable than HS/LS FIM (Fig. \ref{mag_Udep}(c)), 
and LS FM is more stable than LS AFM (Fig. \ref{mag_Udep}(e)).

\subsubsection{HS vs LS with fixed magnetic order: $U$ dependence}
\label{sec:spin_states}

\begin{figure}
\begin{center}
\includegraphics[width=0.5\textwidth, angle=0]{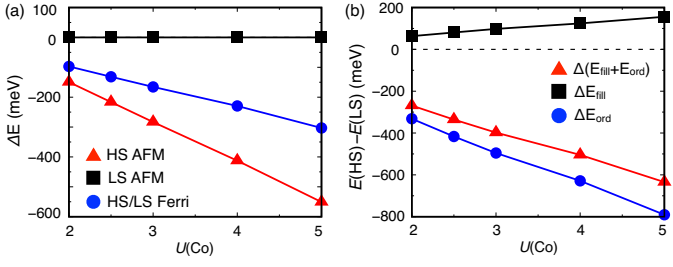}
\caption{ $U_{\textnormal{Co}}$ dependence of the total energies for  AFM states of (LTO)$_1$+(LCO)$_1$ superlattices.  
(a) Total energies per Co of different magnetic configurations (the total energy of the LS state is chosen as the zero of energy).  (b) Spectral decomposition of the $+U$ energy contribution (see text).
}
\label{HS_LS_Udep}
\end{center}
\end{figure}

Interestingly, the above energy differences strongly depend on $U_{\textnormal{Co}}$. 
Specifically, when the magnetic ordering is fixed to AFM, the energy difference between different spin states of Co
increases monotonically as a function of $U_{\textnormal{Co}}$ as shown in Fig.~\ref{HS_LS_Udep}(a).
To explain this strong dependence, we analyze $E$[HS AFM]$-E$[LS AFM] by rewriting the DFT+$U$ energy as a spectral decomposition \cite{Isaacs2016035120,Isaacs2017045141}
\begin{equation}
\begin{aligned}
E =E_{\textnormal{DFT}+U} &= E_{\textnormal{DFT}}  + E_U - E_{\textnormal{dc}}  \\
                                        &= E_{\textnormal{DFT}} + E_{\textnormal{fill}} + E_{\textnormal{ord}} 
\end{aligned}
\label{eq:EDFTUspectral}
\end{equation}
where
\begin{equation}
\begin{aligned}
E_{\textnormal{fill}} &= U(2l+1)\mu(1-\mu), \ \ \ \ E_{\textnormal{ord}}= -U(2l+1) \sigma^2 .
\end{aligned}
\label{eq:EfillEord}
\end{equation}
Here $l=2$ is the angular momentum of the atomic-like states for the DFT+$U$ treatment, and $\mu$ and $\sigma$ are the 
mean value and standard deviation of the eigenvalues of the single particle density matrix 
of the manifold of atomic-like orbtials (here Co $3d$ states). 
Within this spectral decomposition, the combined interaction and double counting
energy simply depends on the two numbers $\mu$ and $\sigma$ extracted from the local single-particle density matrix.

We define $\Delta E_{\textnormal{fill}}^{\rm AFM}$ and $\Delta E_{\textnormal{ord}}^{\rm AFM}$
\begin{equation}
\begin{aligned}
\Delta E_{\textnormal{fill}}^\textnormal{AFM} = E_{\textnormal{fill}}^\textnormal{AFM}[{\rm HS}] - E_{\textnormal{fill}}^\textnormal{AFM}[{\rm LS}]  \\
\Delta E_{\textnormal{ord}}^\textnormal{AFM} = E_{\textnormal{ord}}^\textnormal{AFM}[{\rm HS}] - E_{\textnormal{ord}}^\textnormal{AFM}[{\rm LS}]
\end{aligned}
\end{equation}
and plot them versus $U_{\textnormal{Co}}$ in Fig.~ \ref{HS_LS_Udep}(b).  While the change of $\Delta E_{\textnormal{fill}}^{\rm AFM}$ as a function of $U_{\textnormal{Co}}$ is relatively weak, $\Delta E_{\textnormal{ord}}^{\rm AFM}$ is the dominating contribution, similar to other prior cases \cite{Isaacs2016035120,Isaacs2017045141}.
Therefore, the $U_{\textnormal{Co}}$ behavior is dictated largely by the difference in $\sigma^2$ of the two respective spin configurations.  In other words, the orbital occupancy differentiation (i.e., orbital ordering) is more enhanced in the HS state than in the LS state as $U_{\textnormal{Co}}$ increases, and this enlarges the energy difference between them.

\subsubsection{FM vs AFM for fixed Co spin state: $U_{\textnormal{Co}}$ dependence}
\label{sec:mag_stability}

\begin{figure}
\begin{center}
\includegraphics[width=0.5\textwidth, angle=0]{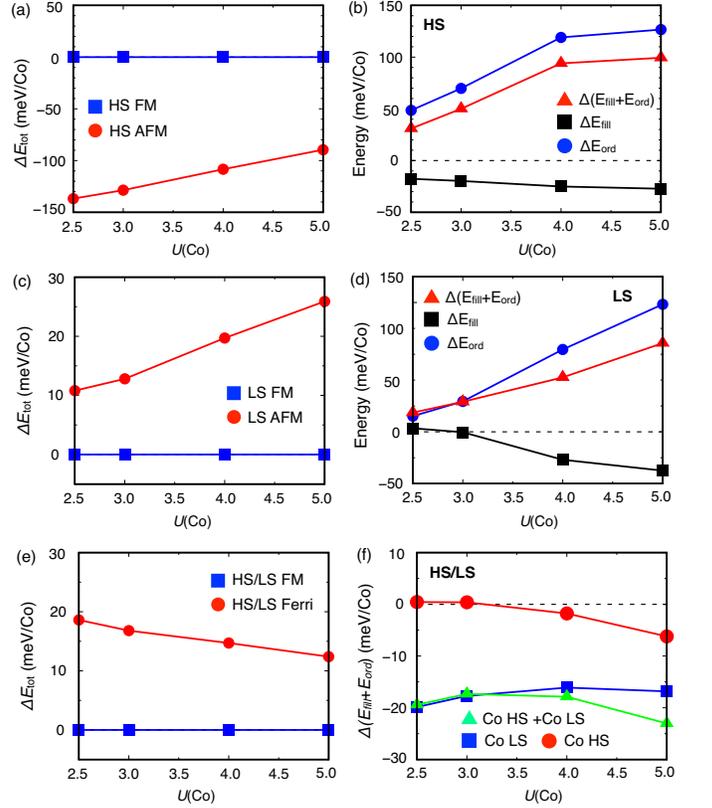}
\caption{  $U_{\textnormal{Co}}$ dependence of energies for different magnetic states of (LTO)$_1$+(LCO)$_1$.
(a) energies of HS FM and HS AFM, where the energy of HS FM is set to be zero.
(c) $\Delta E_{\textnormal{fill}}[\textnormal{HS}]$,  $\Delta E_{\textnormal{ord}}[\textnormal{HS}]$, and 
$\Delta E_{\textnormal{fill}}[\textnormal{HS}]+\Delta E_{\textnormal{ord}}[\textnormal{HS}]$.
(c) Energies of LS FM and LS AFM, where the energy of LS FM is set to be zero.
(c) $\Delta E_{\textnormal{fill}}[\textnormal{LS}]$,  $\Delta E_{\textnormal{ord}}[\textnormal{LS}]$, and 
$\Delta E_{\textnormal{fill}}[\textnormal{LS}]+\Delta E_{\textnormal{ord}}[\textnormal{LS}]$.
(e) Energies of HS/LS FM and HS/LS FIM, where the energy of HS/LS FM is set to be zero.
 (f) $\Delta E_{\textnormal{fill}}[\textnormal{HS/LS}]$,  
$\Delta E_{\textnormal{ord}}[\textnormal{HS/LS}]$, and 
$\Delta E_{\textnormal{fill}}[\textnormal{HS/LS}]+\Delta E_{\textnormal{ord}}[\textnormal{HS/LS}]$.
The energy of the FM phase is set to  zero.
}
\label{mag_Udep}
\end{center}
\end{figure}

Now we consider the energy difference between two magnetic configurations for fixed Co spin state.
Figs.~\ref{mag_Udep}(a) and (c) show that the energy difference between FM and AFM phases.
While the energy difference between FM and AFM decreases as a function of $U_{\textnormal{Co}}$ for the HS state, this energy difference increases as a function of $U_{\textnormal{Co}}$ for the LS state which is the opposite trend.
To explain these trends, we do a similar $U_{\textnormal{Co}}$-dependent spectral analysis for $\Delta E$[HS]$=E^{\rm AFM}$[HS]$-E^{\rm FM}$[HS] (Figs. \ref{mag_Udep}(a) and (b)), 
$\Delta E$[LS]$=E^{\rm AFM}$[LS]$-E^{\rm FM}$[LS] (Figs. \ref{mag_Udep}(c) and (d)), 
and $\Delta E$[HS/LS]$=E^{\rm FIM}$[HS/LS]$-E^{\rm FM}$[HS/LS] (Figs. \ref{mag_Udep}(e) and (f)). $\Delta E_{\textnormal{fill}}$ and $\Delta E_{\textnormal{ord}}$ are defined in a similar way as above: e.g.,
\begin{equation}
\begin{aligned}
\Delta E_{\textnormal{fill}}[\textnormal{HS}] = E_{\textnormal{fill}}^{\rm AFM}[\textnormal{HS}] - E_{\textnormal{fill}}^{\rm FM}[\textnormal{HS}]  \\
\Delta E_{\textnormal{ord}}[\textnormal{HS}] = E_{\textnormal{ord}}^{\rm AFM}[\textnormal{HS}]- E_{\textnormal{ord}}^{\rm FM}[\textnormal{HS}]\,.
\end{aligned}
\end{equation}

Looking at the pair of Figs.~\ref{mag_Udep}(a,b) and the pair Figs.~\ref{mag_Udep}(a,b), the situation is very similar to that in the previous subsection: in both cases $\Delta E_{\textnormal{fill}}$[HS] is almost constant versus $U_{\textnormal{Co}}$ and relatively small in size while $\Delta E_{\textnormal{ord}}$[HS]  dominates the behavior in that its increase with increasing $U_{\textnormal{Co}}$ explains the positive slope of $\Delta E$[HS] or $\Delta E$[LS].

We also considered the intermediate spin HS/LS FM and AFM (really ferrimagnetic as the moments are unequal) configurations where the unit cell contains one LS and one HS Co cation.  
As presented in Fig.~\ref{mag_Udep}(e), $\Delta E$[HS/LS] varies by about 10 meV as $U_{\textnormal{Co}}$ changes from 2.5 to 5 eV. 
$\Delta (E_{\textnormal{fill}}+E_{\textnormal{ord}})$[HS/LS] for the HS Co and the LS Co (see Fig. \ref{mag_Udep}(f)) show opposing behaviors versus  $U_{\textnormal{Co}}$ so their sum is relatively constant .
As a result, the averaged value of $\Delta( E_{\textnormal{fill}}+E_{\textnormal{ord}})$ at $U=2.5$ eV and $U=5$ eV
are only differ by 12 meV.

\subsubsection{Origin of the magnetic stabilities}
\label{sec:mag_origin}

The above numerical observations of the relative magnetic stabilities require  explanations.  Here, we will explain the origin of the magnetic energetics of (LTO)$_1$+(LCO)$_1$ using an energy band picture (i.e., molecular orbital theory) for three different spin configurations: 
HS, HS/LS, and LS.

\begin{figure}
\begin{center}
\includegraphics[width=0.5\textwidth, angle=0]{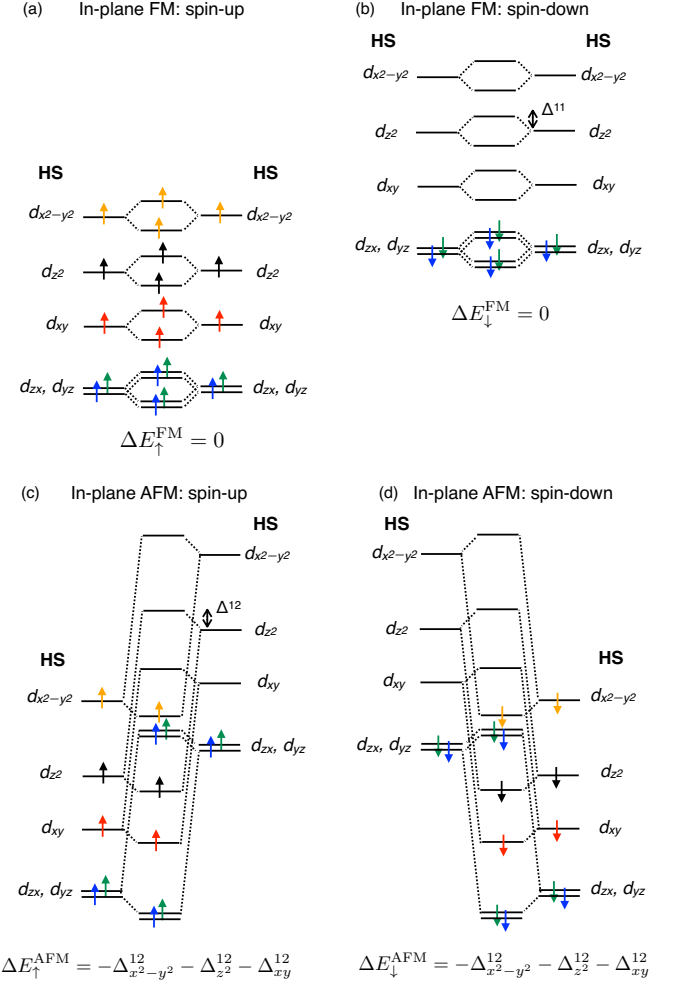}
\caption{  Schematic molecular orbital view of Co 3$d$ energy level diagrams and in-plane magnetic interactions between two neighboring HS Co$^{2+}$ cations. (a) and (b) show two neighboring Co with FM alignment, while (c) and (d) show two neighoring Co with AFM alignment.  Solid horizontal lines indicate energy levels; dashed lines indicate the effect of interactions; arrows indicate electron filling colored by orbital type. }
\label{HS-HS_ip}
\end{center}
\end{figure}

Fig.~\ref{HS-HS_ip} shows the schematic energy diagram of for two in-plane HS Co neighbors with FM and AFM relative spin alignment.  The interaction lines in this diagram do not indicate direct Co 3d-Co 3d hybridization but instead the interaction mediated through the intervening O $2p$ states.  The energetic position of the atomic-like energy levels are drawn based on analysis of projected densities of states diagrams such as Fig.~\ref{HS_LS_pdos}.  

The spin-up majority bands of HS FM are fully occupied, thus the Co-Co interactions do not lead to energetic stabilization as per Fig.~\ref{HS-HS_ip}(a) (i.e., both bonding and anti-bonding states are equally occupied).  We note that the our Co cations have strong orbital degeneracy breaking whereby the $t_{2g}$ $d_{yz}$/$d_{zx}$ bands are lower in energy than the $d_{xy}$ band is empty: in the minority-spin channel, the $d_{yz}$/$d_{zx}$ bands are filled while the $d_{xy}$ is empty. As discussed in Sec.~\ref{sec:LTO1+LCO1}, this splitting results a Mott-like insulating state. In the HS FM case, the minority-spin bands also do not lead to energetic stabilization: the $d_{yz}$($d_{zx}$) spin-down on one Co only interact with the  $d_{yz}$($d_{zx}$) spin-down states on the neighboring Co, and as both are filled, there is no net lowering of energy as the bonding and anti-bonding states are again fully occupied (Fig.~\ref{HS-HS_ip}(b)).  Thus the energy lowering due to Co-Co interactions for HS FM is zero, $\Delta E^{\rm FM}[{\rm HS}]=0$.

On the other hand, HS AFM Co-Co superexchange interactions lead to energetic stabilization (Figs.~\ref{HS-HS_ip}(c,d)).
While one Co has 5 spin-up and 2 spin-down electrons, the other Co has 2 spin-up and 5 spin-down electrons:
$d_{z^2}$, $d_{yz}$ and $d_{zx}$  are fully occupied for both Co ions leading to no energy lowering; but the electrons in the majority-spin
$d_{x^2-y^2}$ and $d_{xy}$ states interact with empty counterparts on the other Co which leads to energy lowering (only bonding states are filled) as shown in Figs.~\ref{HS-HS_ip}(c) and (d).
If we define the strength of the Co-Co interaction for orbital $d$ on one Co with the same orbital on the neighbor for the FM case as $-\Delta^{11}_{d}$ (see Fig.~\ref{HS-HS_ip}(b)) and for the AFM case as $-\Delta^{12}_{d}$ (see Fig.~\ref{HS-HS_ip}(c)), respectively,
then the energy lowering of the spin-up channel in the AFM case is 
\begin{equation}
\begin{aligned}
\Delta E^{\textnormal{AFM}}_{\uparrow}[{\rm HS}] = - \Delta^{12}_{d_{x^2-y^2}} - \Delta^{12}_{d_{z^2}} - \Delta^{12}_{d_{xy}} .
\end{aligned}
\end{equation}
The spin-down channel has same amount of the energy lowering.
Therefore, within this model, the energy difference between HS FM and HS AFM can be written as
\begin{equation}
\begin{aligned}
\Delta E[\textnormal{HS}] 
&= \Delta E^{\textnormal{AFM}}[{\rm HS}] - \Delta E^{\textnormal{FM}}[{\rm HS}] \\
&= -2\left( \Delta^{12}_{d_{x^2-y^2}}+\Delta^{12}_{d_{z^2}}+\Delta^{12}_{d_{xy}} \right)\,,
\end{aligned}
\end{equation}
which is negative as the $\Delta^{ij}_d$ are positive by our convention.  Hence, this analysis provides a simple explanation of the stability of HS AFM compared to HS FM.


\begin{figure}
\begin{center}
\includegraphics[width=0.5\textwidth, angle=0]{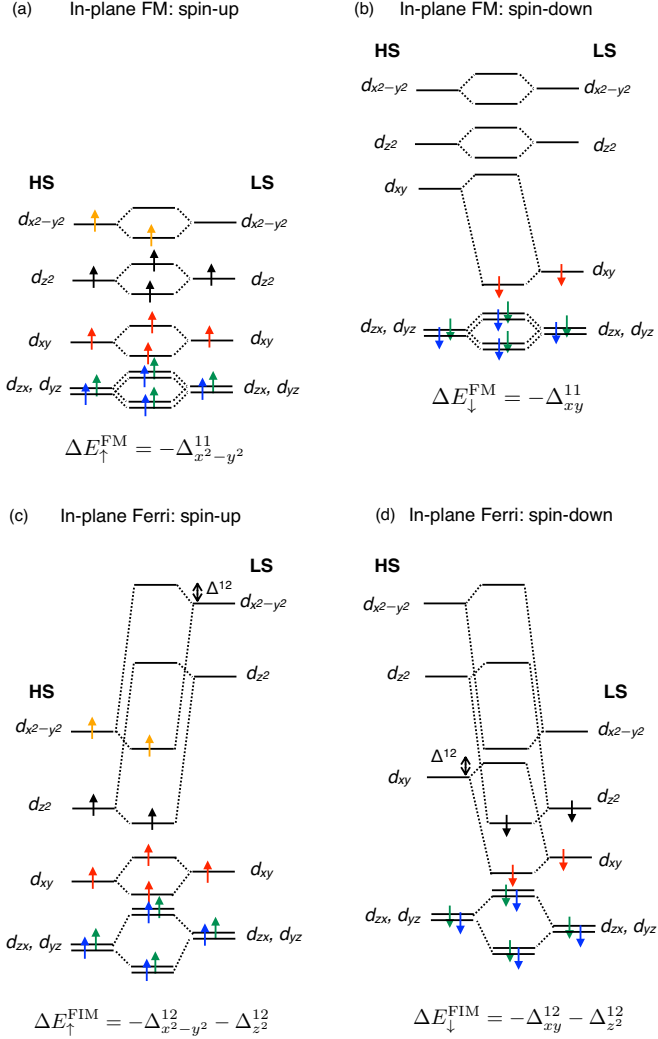}
\caption{  Schematic diagram of the in-plane magnetic interaction between HS Co and LS Co.}
\label{HS/LS_ip}
\end{center}
\end{figure}

Figs.~\ref{HS/LS_ip}(a,b) presents the in-plane interactions between a HS Co and its neighboring LS Co for FM and FIM relative spin alignments.
While the majority spin-up $d$ bands of the HS Co are fully occupied, the $d_{x^2-y^2}$ spin-up band of the LS Co is empty 
(Fig. \ref{HS/LS_ip}(a)) which leads to an energy lowering in the spin-up channel of  $-\Delta^{11}_{d_{x^2-y^2}}$.
In the minority spin-down channel, the $d_{xy}$ band of the LS Co is occupied while the $d_{xy}$ of the HS Co empty, creating a lowering of energy by $-\Delta^{11}_{d_{xy}}$. Thus, the total energy lowering for FM ordering of HS and LS neighboring Co is
\begin{equation}
\begin{aligned}
\Delta E^{\textnormal{FM}}[{\rm HS/LS}] = - \Delta^{11}_{d_{x^2-y^2}} - \Delta^{11}_{d_{xy}} .
\end{aligned}
\end{equation}

The case of ferrimagnetic (FIM) ordering between a HS Co and LS Co is  shown in Figs.~\ref{HS/LS_ip}(c,d). For the spin-up channel, the HS $d_{x^2-y^2}$ and $d_{z^2}$ are occupied but empty for the LS Co, leading to an energy lowering of $- \Delta^{12}_{d_{x^2-y^2}} - \Delta^{12}_{d_{z^2}}$; a similar lowering happens for the low-spin channel due to $d_{xy}$ and $d_{z^2}$.  The total energetic stabilization for this FIM HS--LS case 
\begin{multline}
\Delta E^{\textnormal{FIM}}[{\rm HS/LS}] = -2\Delta^{12}_{d_{z^2}} 
- \Delta^{12}_{d_{x^2-y^2}} - \Delta^{12}_{d_{xy}}\,.
\end{multline}

Hence, the energy difference between HS/LS FM and HS/LS FIM is thus
\begin{multline}
\Delta E[\textnormal{HS/LS}] 
 = \Delta E^{\textnormal{FM}}[{\rm HS/LS}] - \Delta E^{\textnormal{FIM}}[{\rm HS/LS}] \\
= - \left[ \Delta^{11}_{d_{x^2-y^2}} - \Delta^{12}_{d_{x^2-y^2}} \right] - \left[ \Delta^{11}_{d_{xy}} - \Delta^{12}_{d_{xy}}\right]\\ + 2\Delta^{12}_{d_{z^2}}\,.
\end{multline}
For the $d_{x^2-y^2}$ and $d_{xy}$ contributions, there is some partial cancellation between the $\Delta^{11}$ and $\Delta^{12}$ terms although these contributions should be net negative since FM interactions $\Delta^{11}$ take place between energy degenerate orbitals while the AFM $\Delta^{12}$ are between strongly exchange-split 3d orbitals leading to weaker interactions (i.e., homopolar versus heteropolar bonding in the language of semiconductors).  However, with the addition of the positive contribution $2\Delta^{12}_{d_{z^2}}$, the total is expected to be relatively small in magnitude in agreement the small magnitude of energy differences shown in Fig.~\ref{mag_Udep}(e).  Comparison to data in the figure shows that $\Delta^{11} > \Delta^{12}$ must be true to the extent that it leads to FIM to be higher in energy than FM.


\begin{figure}
\begin{center}
\includegraphics[width=0.5\textwidth, angle=0]{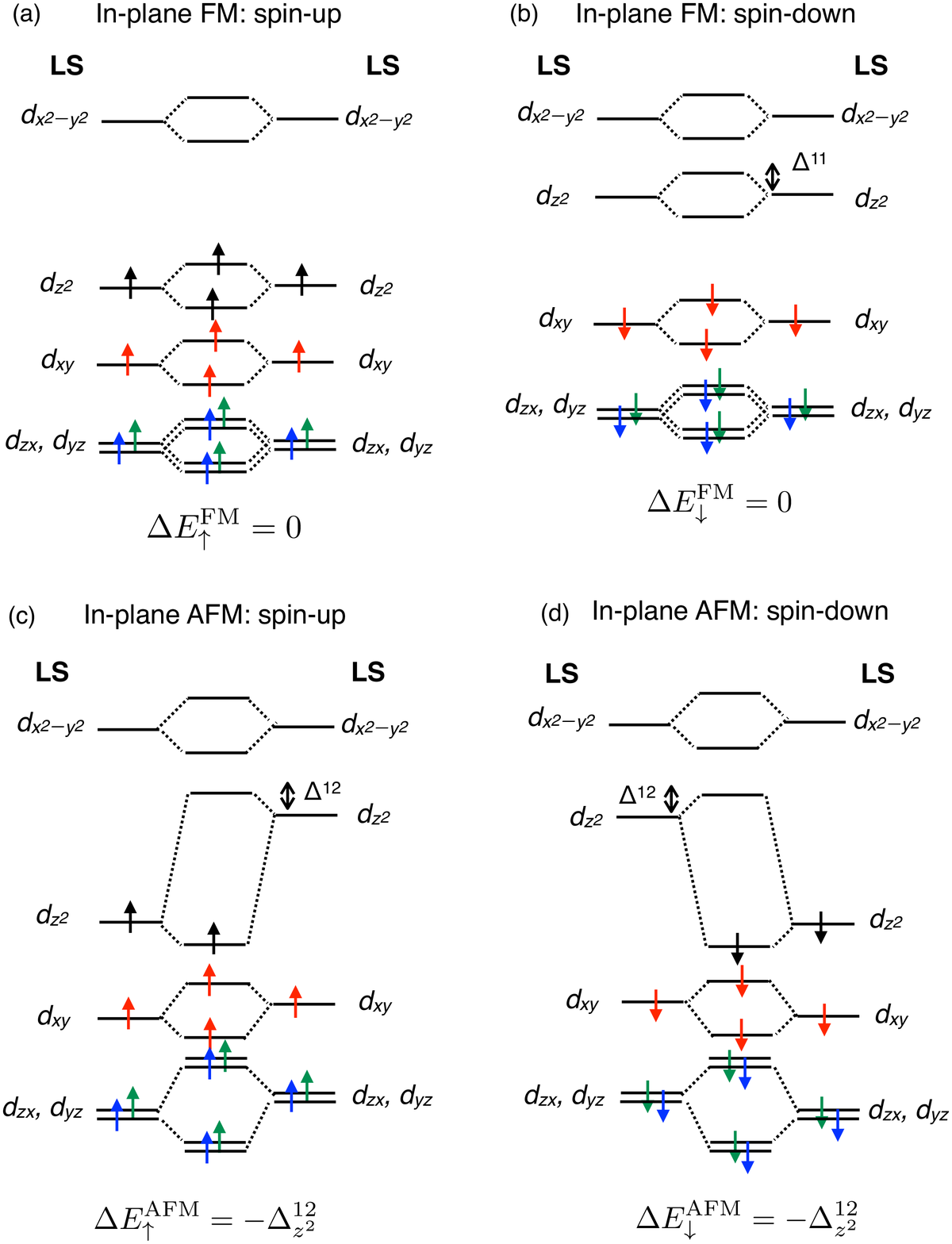}
\caption{  Schematic diagram of the in-plane magnetic interaction between LS Co and LS Co.   }
\label{LS-LS_ip}
\end{center}
\end{figure}

Understanding the stability of the FM state for two LS Co neighbors, however, is more complicated. The simple band interaction picture cannot explain the data in Fig.~\ref{mag_Udep}(e) which show LS FM is more stable than LS AFM: as per Fig.~\ref{LS-LS_ip}, the energy lowering of the FM configuration is zero while the AFM ordering should be lower in energy by $-2\Delta^{12}_{d_{z^2}}$.  Furthermore, standard double exchange for FM configurations is unlikely to be a viable explanation since the system is not metallic (it has an energy gap) due to the large  splitting between $t_{2g}$ and $e_g$ bands.  
Nor can Goodenough-Kanamori (GK) type FM superexchange be the origin of the stability, 
because GK superexchange is maximum if $\angle$TM--O--TM=90$^\circ$ and cannot be applied for $\angle$TM--O--TM=180$^\circ$
\cite{GoodenoughPRB1955,GoodenoughJPCS1958,KanamoriJPCS1959}.
We find that FM is also stable for the insulating undistorted $a^0a^0a^0$ phase, where Co--O--Co angle is 180$^\circ$.

\begin{figure}
\begin{center}
\includegraphics[width=0.5\textwidth, angle=0]{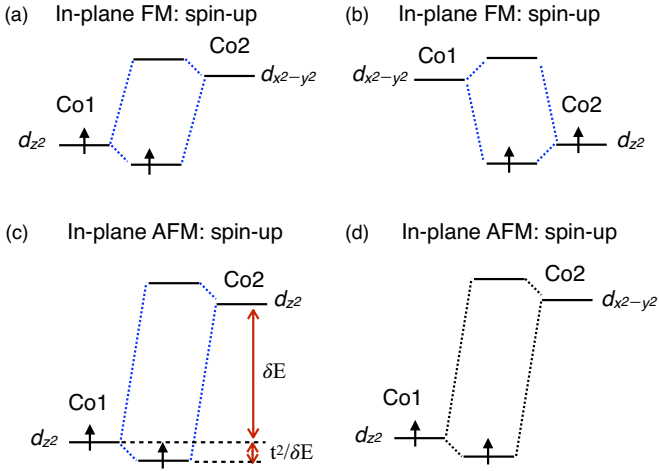}
\caption{  Schematic magnetic interaction model between two neighboring LS Co for (a,b) FM spin order, and (c,d) AFM spin order. Panel (c) shows the  difference in on-site energy $\delta E$ and how the hopping between the two orbitals $t$ leads to an energetic lowering by $t^2/\delta E$ for the bonding state; numerical values are provided in Table~\ref{wannier_parameters}.
}
\label{LS-LS_model}
\end{center}
\end{figure}

In the above analysis, we only considered the interaction between same $d$ orbitals on the neighboring Co, e.g.,
$d_{z^2}$--$d_{z^2}$ or $d_{xy}$--$d_{xy}$, since they are usually the dominant ones.  The model predicted that the LS AFM state is stabilized by $-2\Delta^{12}_{d_{z^2}}$: but will be a very small quantity in this case since the exchange splitting of the $d_{z^2}$ band is larger than 4 eV (see Fig.~\ref{HS_LS_pdos}(b)) and we expect the hopping between the two neighboring $d_{z^2}$ orbitals to be small since they are strongly directional out-of-plane.  Hence, it is likely that interactions between different orbitals on the neighboring Co can play a role in this case.  

To capture the essence of the problem, we consider the ideal $a^0a^0a^0$ structure without octrahedral tilts so the Co--O--Co bonding is along a straight line.  In this case, the only symmetry allowed interactions that can lower the energy for the FM order are $d_{z^2}$--$d_{x^2-y2}$ between neighboring atoms for the majority spin-up channel (Figs.~\ref{LS-LS_model}(a,b)).  For AFM order, the relevant hoppings are $d_{z^2}$--$d_{z^2}$ (small and already discussed above) and $d_{z^2}$--$d_{x^2-y^2}$ in both spin channels as per Figs.~\ref{LS-LS_model}(c,d).

\begin{table}
\begin{ruledtabular}
\begin{center}
\caption{Hopping parameters ($t$) and energy splitting of $d_{z^2}$ and $d_{z^2}$ bands ($\delta E$) as obtained from a Maximally Localize Wannier Function analysis.  }
\label{wannier_parameters}
\renewcommand{\arraystretch}{1.3}
\begin{tabular}{c c c c c}
order	& interaction & $\delta E$ (eV)  & $t$ (eV) & $t^2/\delta E$ (eV)\\
 \hline
 FM	& $d_{z^2} - d_{x^2-y2}$	& 2.04	& $-$0.120	& 7.0$\times10^{-3}$ \\
AFM	& $d_{z^2} - d_{z^2}$		& 3.38	& $-$0.010	& 2.9$\times10^{-5}$ \\ 
AFM	& $d_{z^2} - d_{x^2-y2}$	& 2.98	& $-$0.116 	& 4.5$\times10^{-3}$ \\
\end{tabular}
\end{center}
\end{ruledtabular}
\end{table}

We now create a more quantitative model by extracting on-site energies and hopping parameters for the LS FM and AFM cases by generating Maximally Localize Wannier Functions (MWLF)~\cite{wannier90} that span the energy bands for the $d_{z^2}$ and $d_{x^2-y2}$ character bands near the Fermi level.  These are low-energy or effective Wannier functions that are centered on the Co cations and span the space of the relevant bands (i.e., this is not a p-d model as we don not create any oxygen 2p-like Wannier functions).  Table~\ref{wannier_parameters} displays the key band parameters that describe the hoppings $t$ between neighboring orbitals and their on-site energy differences $\delta E$.  The energy lowering for an interacting pair of neighboring orbitals with a single electron between them is estimated by perturbation theory to be $\Delta=-t^2/\delta E$. Table~\ref{wannier_parameters} shows that (a) the AFM $\Delta^{12}_{d_{z^2}}$  is indeed very weak because the hopping element $t$ is so small, and (b) the FM $d_{z^2}$--$d_{x^2-y^2}$ interaction is more stabilizing due to the smaller energy splitting $\delta E$.  We conclude that the LS FM state is more stable than LS AFM, and the stabilization is driven by more favorable $d_{z^2}$--$d_{x^2-y^2}$ interactions due to the smaller $d_{z^2}$--$d_{x^2-y^2}$ energy splitting in the FM case between Co neighbors.

\subsubsection{Effect of strain and exchange-correlation functional}
\label{sec:strain}

\begin{figure}
\begin{center}
\includegraphics[width=0.5\textwidth, angle=0]{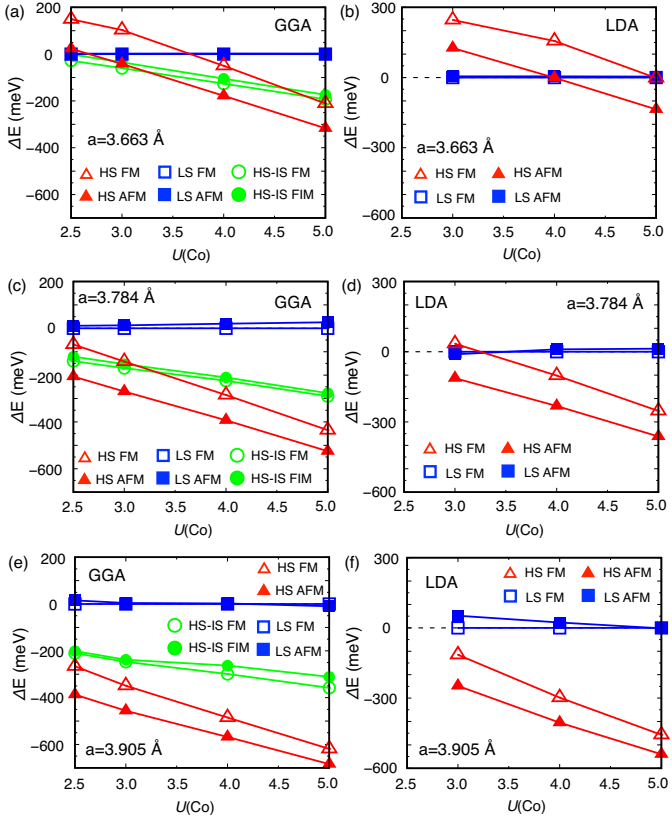}
\caption{  $U_{\textnormal{Co}}$ dependence of the energies of different magnetic states of (LTO)$_1$+(LCO)$_1$ 
for different in-plane lattice parameters (3.663 \AA{} for (a,b), 3.784 \AA{} for (c,d), and 3.905 \AA{} for (e,f)) and two different exchange-correlation functionals ((a,c,e) for GGA+$U$ and (b,d,f) for LDA+$U$).  Energies of LS FM phase is set to zero.
}
\label{strain_Udep}
\end{center}
\end{figure}

Given that the spin, orbital, and lattice degrees of freedom strongly interact in a transition metal octahedron, 
the electronic and magnetic properties of Co in (LTO)$)1$+(LCO)$_1$  depend on strain strongly.
To study this, we consider the three different in-plane lattice parameters discussed above ($a=3.663$ \AA, 3.784 \AA, and 3.905 \AA).
Within GGA+$U$ for $U_{\textnormal{Co}}=3$ eV, these three lattice parameters
corresponds to 3.7\%  compressive, 0.6\%  compressive, and and 2.6\%  tensile strain.


In Fig. \ref{strain_Udep}, we summarize the strain-dependent stabilities of different spin states and 
magnetic orderings of Co within both GGA+$U$ and LDA+$U$.
Interestingly, as shown in Figs.~\ref{strain_Udep}(a,c,e),
the HS phases are stabilized greatly as the lattice parameter increases: the
LS phase is stabilized by compressive strain, whereas the HS phase is stabilized by tensile strain.
In addition, similar to the case of $a=3.784$ \AA{} discussed in detail above in Sec.~\ref{sec:spin_states},
HS FM and HS AFM phases are stabilized as $U_{\textnormal{Co}}$ increases 
(see Figs.~\ref{strain_Udep}(a,c,e)).  Various combinations of strain and $U_{\textnormal{Co}}$ can drive magnetic transitions for the Co$^{2+}$: LS FM can be stabilized over the HS AFM for compressive strain and low  $U_{\textnormal{Co}}$ and an intermediate spin (HS/LS combination) state can become the ground state (e.g., $U_{\textnormal{Co}}=2.5$ eV and $a=3.663$ \AA).  However, over broad ranges of parameters, it is the HS AFM that is the ground state.  To the best of our knowledge, HS to LS or FM to AFM transitions have not been suggested or observed for Co$^{2+}$ in bulk CoO, and this makes the LCO+LTO superlattice a potential test bed for modifying the spin state and magnetic interactions of Co$^{2+}$ ions.

We find that the $U_{\textnormal{Co}}$ dependence of the energy difference between HS and LS phases originates from the orbital occupancy differentiation energy $E_{\textnormal{ord}}$ (Eq.~\ref{eq:EfillEord}) much like what was shown in Fig.~\ref{HS_LS_Udep}.  On the other hand, we do not find that the relative stabilization of the LS phase for compressive strain is connected to the atomic-like $+U$ contributions to the total energy $E_{\textnormal{ord}}$ nor $E_{\textnormal{fill}}$ but must instead originate from the underlying DFT energy $E_{DFT}$ that encodes the electronic dispersion and hopping contributions.  The idea that the DFT part of the total energy controls the relative stability of the LS phase means that its stability should strongly depend on the choice of DFT functional. This is borne out by  Fig.~\ref{strain_Udep} which compares the 
 GGA and LDA functionals.  The trends versus $U_{\textnormal{Co}}$ and strain are very similar between the two functionals, but the energy of the HS phases is   shifted upwards compared to the LS phases as we switch from GGA to LDA.

\subsubsection{Effect of doping}
\label{sec:doping}

\begin{figure}  
\begin{center}
\includegraphics[width=0.48\textwidth, angle=0]{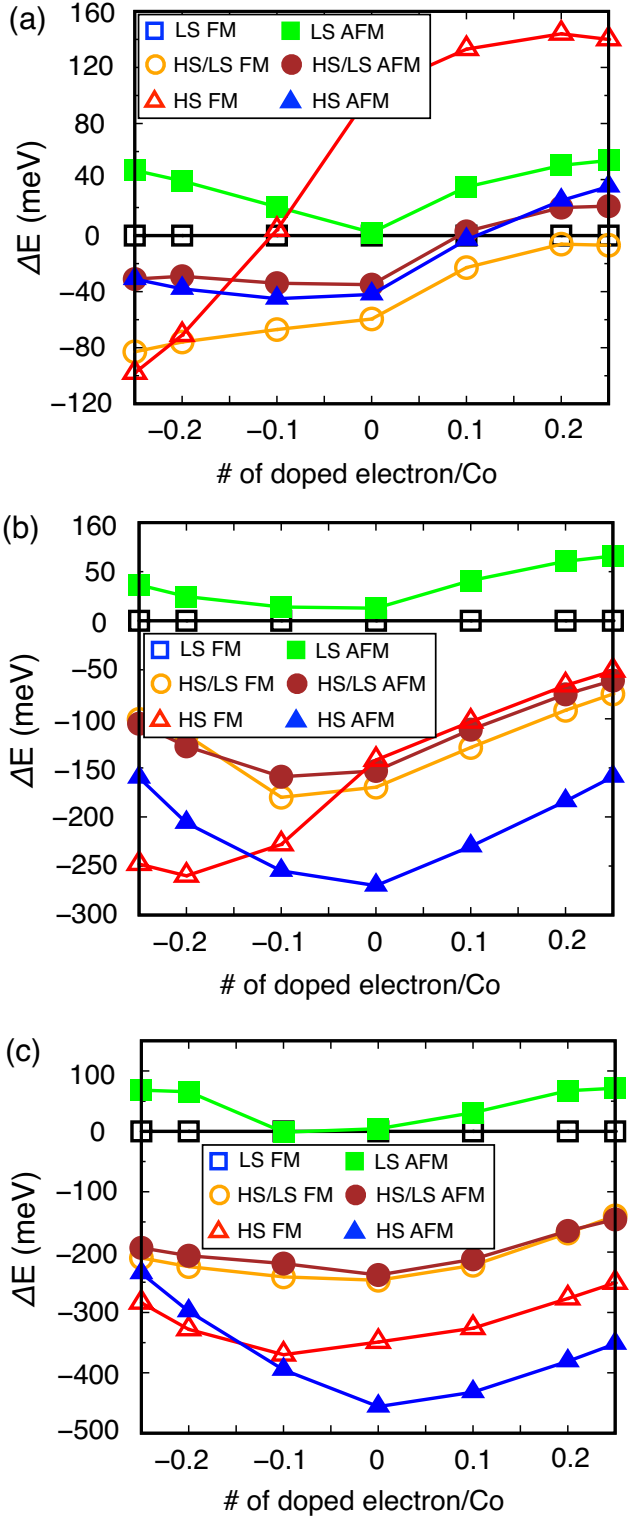}
\caption{  Magnetic stabilities of electron- and hole-doped (LTO)$_1$+(LCO)$_1$ superlattices,
with (a) $a=3.663$\AA, (b) $a=3.784$ \AA, and (c) $a=3.905$ \AA.
The energy  of the LS FM phase is set to  zero.  }
\label{1-1_dop}
\end{center}
\end{figure}

Electron or hole doping is a general and powerful way to change the magnetic stability of the transition metal oxides such as cuprates \cite{cuprateRMP2006,cuprateRMP2018} and manganites \cite{SalamonRMP2001,HellmanRMP2017}. 
By manually changing total number of electrons in the super cell,
we find that both the spin state and the magnetic ordering can  be changed via electron or hole doping in (LTO)+(LCO) superlattices.
In Fig. \ref{1-1_dop}, we present the energies of six different phases as a function of electron/hole doping.
The spin of each Co ion is in fact different since the added electron or hole  resides on one of the Co ions.  Thus, what is referred to as an antiferromagnetic alignment of Co spins is in fact a ferrimagnetic ordering,
but for simplicity we refer to such phases by their parent undoped phase as AFM in Fig. \ref{1-1_dop}.

Overall, the figure shows that relative energetic stabilities are essentially unchanged by electron doping: the HS/LS FM phase is most stable at $a=3.663$ \AA{} (Fig. \ref{1-1_dop}(a)), 
and the HS AFM phase is most stable at $a=3.784$ \AA{} and $a=3.905$ \AA{} (Figs. \ref{1-1_dop}(b,c)) for all considered electron doping values.  On the other hand, the relative magnetic stabilities are changed dramatically by  hole doping.
Specifically, the energies of the hole-doped HS phases show large changes
(red and blue curves in Fig. \ref{1-1_dop}). 
HS FM phase becomes the ground state if the hole doping per Co is larger than  $\sim$0.2, $\sim$0.13, and $\sim$0.15 for $a=3.663$, 3.784, and 3.905 \AA, respectively.
This result is particularly interesting, because the hole doping via Sr substitution on La sites should be much easier to achieve in LTO-based systems than electron doping.

\begin{figure}  
\begin{center}
\includegraphics[width=0.5\textwidth, angle=0]{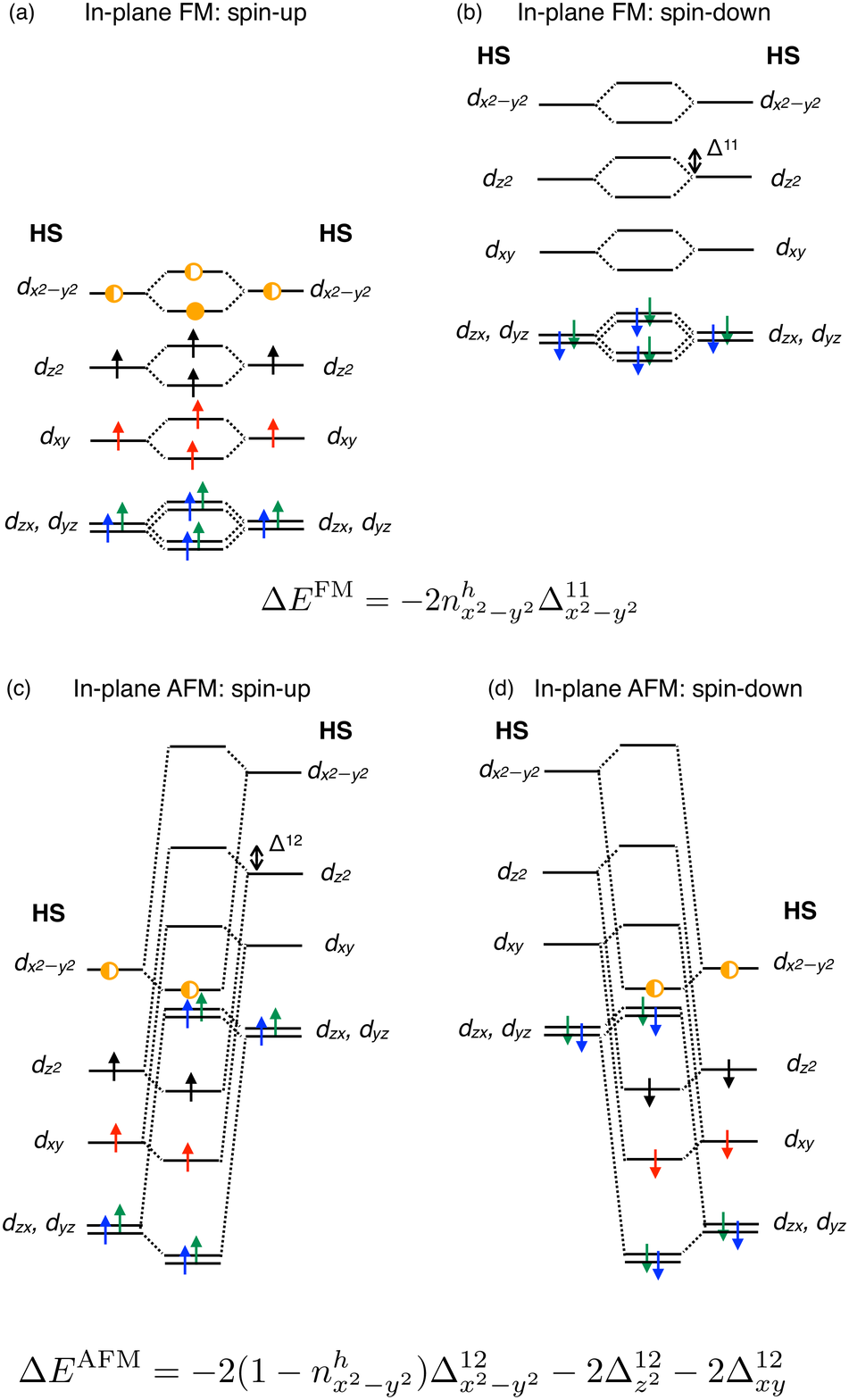}
\caption{  Schematic diagram of the in-plane magnetic interaction between two neighboring HS Co and HS Co for FM (a,b) and AFM (c,d) configurations in hole-doped (LTO)$_1$+(LCO)$_1$.   }
\label{HS-HS_dop}
\end{center}
\end{figure}

We find that both the electron-doped and hole-doped HS phases become metallic.
Doped electrons partially occupy the minority spin-down $d_{xy}$ states, and doped holes resides on majority spin-up $d_{z^2}$ and $d_{x^2-y^2}$ states.
Given the metallicity of the hole-doped state, the stability of the FM ordering is most likely due to the double exchange mechanism.  We use the schematic band interaction diagrams of in Fig. \ref{HS-HS_dop} to explain this stabilization mechanism. As per Fig. \ref{HS-HS_dop}(a), the energy lowering of FM HS ordering is nonzero because of the holes in majority spin-up $d_{z^2}$ and $d_{x^2-y^2}$ bands. Defining the number of holes as $n^{h}_{d_{z^2}}$ and $n^{h}_{d_{x^2-y^2}}$, the energy lowering of the FM ordering  is 
\begin{equation}
\begin{aligned}
\Delta E^{\textnormal{FM}} = 
-2n^{h}_{x^2-y^2} \Delta^{11}_{x^2-y^2}  , 
\end{aligned}
\end{equation}
where the factor of two is due to the holes on both neighboring Co migrating to the shared anti-bonding $d_{x^2-y^2}$ state.  The energy lowering of the AFM ordering (strictly speaking, it is ferrimagnetic as discussed above) is 
\begin{equation}
\begin{aligned}
\Delta E^{\textnormal{AFM}} = 
&-\left( 2- 2n^{h}_{x^2-y^2} \right) \Delta^{12}_{x^2-y^2} \\
&-2 \Delta^{12}_{z^2}
-2\Delta^{12}_{xy} , 
\end{aligned}
\end{equation}

The energy difference between the two phases is thus
\begin{equation}
\begin{aligned}
\Delta E^{\textnormal{FM}} - \Delta E^{\textnormal{AFM}}   &= -2 n^{h}_{x^2-y^2} \left( \Delta^{11}_{x^2-y^2} + \Delta^{12}_{x^2-y^2} \right) \\
&+2 \left(  \Delta^{12}_{xy} +\Delta^{12}_{x^2-y^2} +\Delta^{12}_{z^2} \right)\,.
\end{aligned}
\end{equation}
This energy difference can change sign with increasing hole doping $n^{h}_{d_{x^2-y^2}}$.  Furthermore, since $\Delta^{11}_d$ energies are much larger than $\Delta^{12}_d$ energies, the value of $n^{h}_{d_{x^2-y^2}}$ at which it changes sign should be small in agreement with our above numerical findings.

\subsection{(LaTiO$_3$)$_2$+(LaCoO$_3$)$_2$ superlattice}
\label{sec:LTO2+LCO2}

\begin{figure*}
\begin{center}
\includegraphics[width=0.8\textwidth, angle=0]{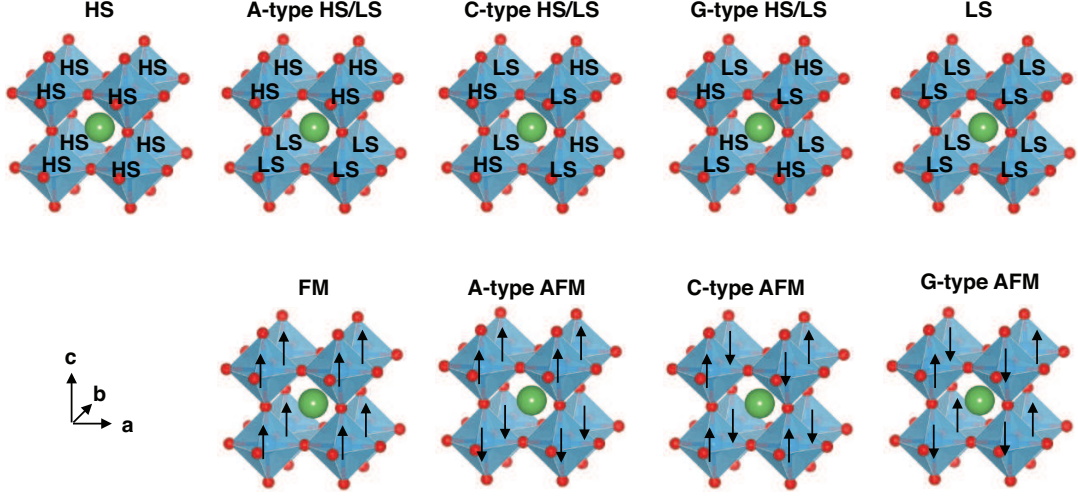}
\caption{  5 possible HS/LS orderings, and 4 possible magnetic orderings; total possible configurations are 5$\times$4=20.   
}
\label{2-2_configs}
\end{center}
\end{figure*}

\begin{figure}
\begin{center}
\includegraphics[width=0.49\textwidth, angle=0]{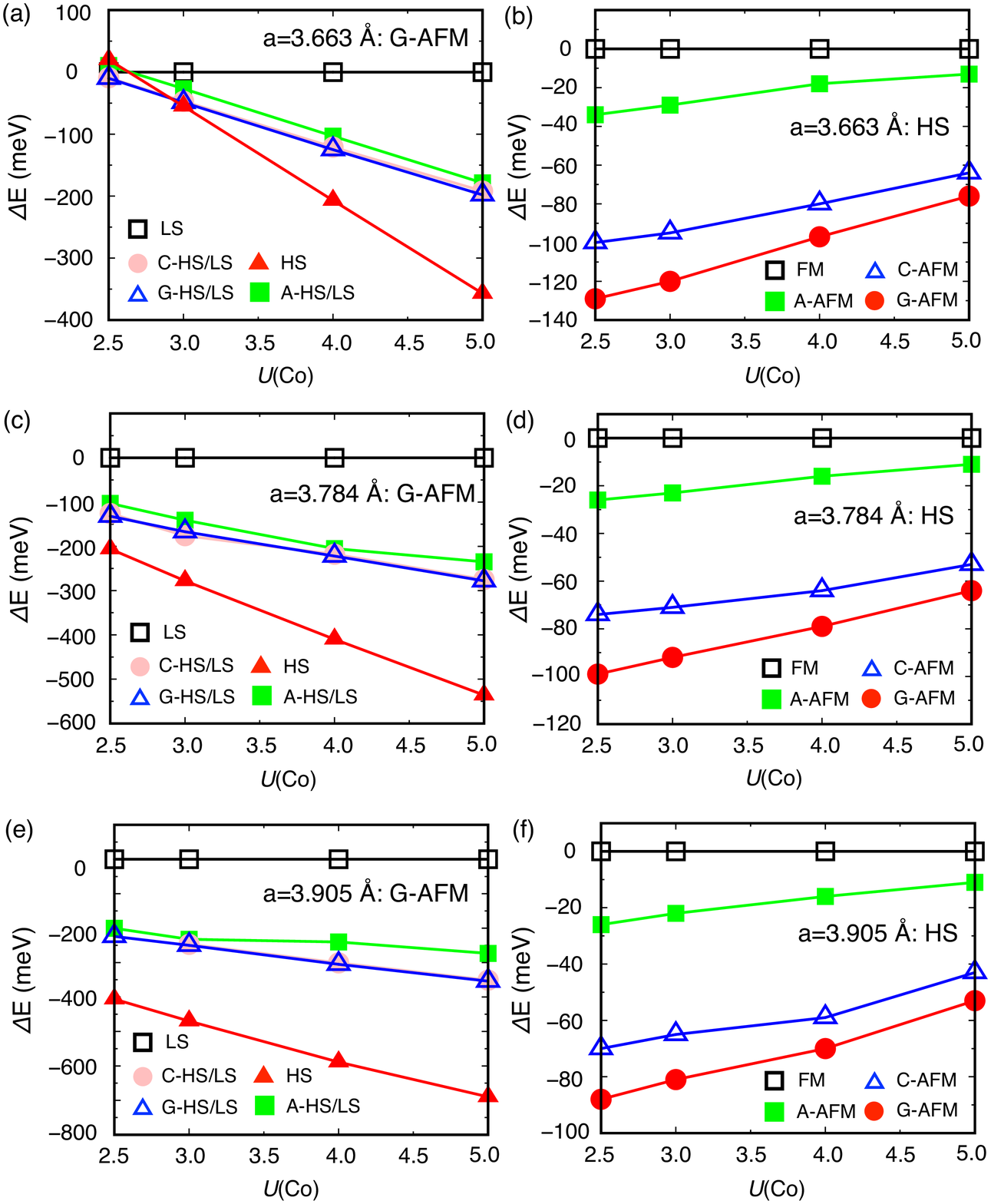}
\caption{  $U$ dependence of energies for different spin configurations of (LTO)$_2$+(LCO)$_2$ 
with different in-plane lattice parameters. Energy of LS FM is set to be zero.
}
\label{2-2_Udep}
\end{center}
\end{figure}

We now turn to describe the electronic properties of (LCO)$_2$+(LTO)$_2$ superlattices.  Our computational supercells contain 4 Co and 4 Ti.  Since there is a great deal of shared physics between the 1+1 and 2+2 superlattices, we will only describe the main physical properties of the 2+2 superlattices while highlighting key differences: our overall approach and methods of analysis are idential for the two superlattices.

The Co$^{2+}$ cations in the (LCO)$_2$+(LTO)$_2$ superlattice can be HS or LS, similar to (LCO)$_1$+(LTO)$_1$, so that there are 5 possible  configurations HS/LS ordering in the supercell as presented in top row of Fig.~\ref{2-2_configs}: all HS order (shortened to HS), A-type HS/LS order with ordering wave vector $\bm{q_\textnormal{sc}} = \left(0,0,\frac{1}{2}\right)$ (A-HS/LS), 
C-type HS/LS order with $\bm{q_\textnormal{sc}} = \left(\frac{1}{2},\frac{1}{2},0 \right)$ (C-HS/LS), 
G-type HS/LS order with $\bm{q_\textnormal{sc}} = \left(\frac{1}{2},\frac{1}{2},\frac{1}{2}\right)$ (G-HS/LS),
and all LS order (LS).
Separately, there are 4 possible orderings patterns for the Co magnetic moments: FM, A-type AFM, C-type AFM, and G-type AFM (see second row of Fig.~\ref{2-2_configs}).  In total, we investigate all $5\times4=20$ configurations.

We summarize the $U_{\textnormal{Co}}$-dependent properties of the most relevant low-energy configurations in Fig.~\ref{2-2_Udep}.  For most choices of parameters, the full set of $U_{\textnormal{Co}}$-dependent data show that HS G-AFM is the ground state, except for the case of high compressive strain and small $U_{\textnormal{Co}}$ ($a$=3.663 \AA{} and $U_{\textnormal{Co}}$=2.5 eV)
where we find that G-HS/LS A-AFM phase is more stable (to avoid excessive clutter, Fig.~\ref{2-2_Udep}(a) only shows data for G-AFM magnetic ordering).
In general, the energies of the A-HS/LS, C-HS/LS, G-HS/LS phases lie between those of the LS and HS phases.
For a fixed magnetic pattern (Fig.~\ref{2-2_Udep}(a,c,e)), 
the energy difference ($\Delta E$) between LS and HS phases is enhanced
as $U_{\textnormal{Co}}$ increases or in-plane lattice parameter $a$ increases which is similar to the behavior of (LTO)$_1$+(LCO)$_1$.
Again, like the behavior of the 1+1 superlattices described in Sec.~\ref{sec:spin_states}, the $U_{\textnormal{Co}}$-dependence of $\Delta E$ for 2+2 superlattices turns out to be explained by the behavior of the ordering energy $\Delta E_{\textnormal{ord}}$ (Eqs.~(\ref{eq:EDFTUspectral},\ref{eq:EfillEord})) between LS and HS phases.

In Figs.~\ref{2-2_Udep}(b,d,f), we compare the energies of different magnetic orderings as a function of $U_{\textnormal{Co}}$ when all then Co are  in the HS state.  The energies in this case obey the relations
\[
E^{\rm FM}[{\rm HS}] > E^\textnormal{A-AFM}[{\rm HS}] > E^\textnormal{C-AFM}[{\rm HS}] > E^\textnormal{G-AFM}[{\rm HS}],
\]
and interestingly this order holds for all $U_{\textnormal{Co}}$ and strain ranges  considered.
\begin{figure} 
\begin{center}
\includegraphics[width=0.49\textwidth, angle=0]{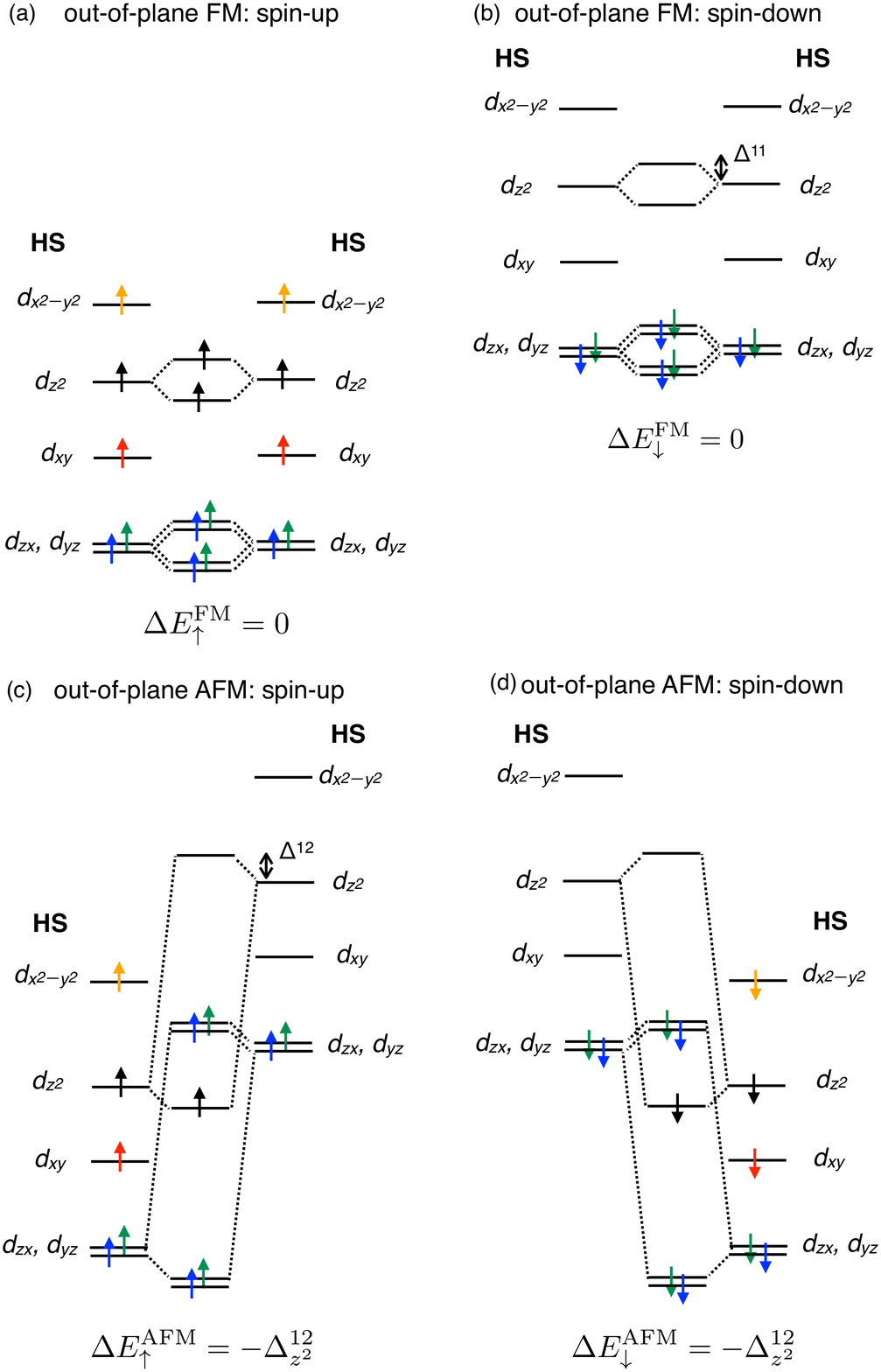}
\caption{  Schematic diagram of the out-of-plane magnetic interaction between HS Co and HS Co.   }
\label{HS-HS_oop}
\end{center}
\end{figure}
As described in Fig.~\ref{HS-HS_ip} and its associated discussion, in-plane magnetic interactions stabilizes AFM spin alignment,
thus $E^\textnormal{FM}[\textnormal{HS}]$ and $E^\textnormal{A-AFM}[\textnormal{HS}]$ are above $E^\textnormal{C-AFM}[\textnormal{HS}]$ and $E^\textnormal{G-AFM}[\textnormal{HS}]$.  

In order to explain why $E^\textnormal{FM}[\textnormal{HS}]> E^\textnormal{A-AFM}[\textnormal{HS}]$ and $E^\textnormal{C-AFM}[\textnormal{HS}]> E^\textnormal{G-AFM}[\textnormal{HS}]$,
i.e., the stability AFM order along the out-of-plane direction,
we examine the out-of-plane magnetic interaction for the HS phase in Fig.~\ref{HS-HS_oop}.
In the out-of-plane direction, only $d_{z^2}$, $d_{zx}$, and $d_{yz}$ on two neighboring Co will interact significantly.
While the FM alignment has no net energy lowering due to out-of-plane interactions (see Figs.~\ref{HS-HS_oop}(a,b)), the
interactions between the $d_{z^2}$ on the two Co lowers the energy by $-2\Delta^{12}_{d_{z^2}}$ for  AFM alignment (Figs.~\ref{HS-HS_oop}(c,d)).
The energy difference between the FM and AFM configurations due to out-of-plane interactions is 
\begin{equation}
\begin{aligned}
\Delta E[\textnormal{HS}] 
=\Delta E^\textnormal{AFM}[\textnormal{HS}]-\Delta E^\textnormal{FM}[\textnormal{HS}]  = -2 \Delta^{12}_{d_{z^2}}\,,
\end{aligned}
\end{equation} 
so that we expect $E^\textnormal{AFM}[\textnormal{HS}] < E^\textnormal{FM}[\textnormal{HS}]$ and $E^\textnormal{G-AFM}[\textnormal{HS}] < E^\textnormal{C-AFM}[\textnormal{HS}]$.

\section{Summary and outlook}

LCO+LTO superlattices show strong electron transfer, significant structural distortion, and robust orbital polarization.  For a wide range of computational parameters, we predict high-spin Co and a checkerboard antiferromagnetic (AFM) order on the Co superlattice (2D checkerboard for the 1+1 superlattice and 3D checkerboard, i.e., G-type, for the 2+2 superlattice).  A detailed analysis of the electronic structure reveals how superexchange interactions control the stability of the magnetic orders.  Finally, hole doping of the superlattices can lead to stabilization of a ferromagnetic ground state of Co$^{2+}$.  This is unexpected and interesting especially for future experiments.  Generally, materials containing Co$^{2+}$ as the sole magnetic cation, such as CoO or Co$_3$O$_4$ (which also contains non-magnetic Co$^{3+}$ \cite{DuttaJPCM2007}), show antiferromagnetic ordering.  A ferrimagnetic compound such as CoFe$_2$O$_4$ contains Fe$^{3+}$ at the $A$ sites and Fe$^{3+}$/Co$^{2+}$ at $B$ sites which are aligned antiferromagnetically \cite{Stein2018}.  Hence, the ability to create a ferromagnet using high-spin Co$^{2+}$ is a new and surprising prediction.

Experimentally, we hope to see verification of the checkerboard AFM order of these superlattices.  Even more exciting will be the experimental realization of ferromagnetism via hole doping through replacement of La by Sr; by careful control of the doping, the FM and AFM order can be made nearly degenerate in energy, potentially leading to interesting magnetic response and new magnetic phases.

Theoretically, more advanced models such as DFT + dynamical mean field theory can be used to verify the predictions made here using GGA+$U$.    Furthermore, understanding the origin of the orbital polarization in these charge-transfer systems is interesting in terms of basic science as well as electronic engineering.  Finally, investigation of thicker LCO layers in the superlattice, i.e., (LTO)$_2$+(LCO)$_n$ for $n\ge3$, should prove  interesting  as well, since the interaction between Co$^{2+}$ at the interface and Co$^{3+}$ at the bulk-like region may lead to other unexpected physical phenomena.

\section{Acknowledgments}
We thank to Sangjae Lee, F. J. Walker, and Charles H. Ahn for helpful discussions. We thank the Yale Center for Research Computing for guidance and use of the research computing infrastructure.  This work also used the Extreme Science and Engineering Discovery Environment (XSEDE), which is supported by National Science Foundation grant number ACI-1548562, by using computer time on the Comet supercomputer as enabled by XSEDE allocation MCA08X007.

\bibliography{myref}

\end{document}